\documentclass[twocolumn,longauth]{aa} %
\usepackage{morefloats}
\usepackage{graphicx}
\usepackage{xcolor,color}
\usepackage{ragged2e}
\usepackage{txfonts}
\usepackage[colorlinks=true, linkcolor=blue, citecolor=blue, urlcolor=blue]{hyperref}
\usepackage{float}

\begin{document}

   \title{Impact of AGN and nuclear star formation on the ISM turbulence of galaxies: Insights from JWST/MIRI spectroscopy}
   \titlerunning{Impact of AGN and SF on the ISM turbulence of galaxies}

   \author{Rogemar A. Riffel\inst{1,2} 
    \and Luis Colina\inst{1}
    \and José Henrique Costa-Souza\inst{2,1} 
    \and Vincenzo Mainieri\inst{3}
    \and Miguel Pereira Santaella\inst{4}
    \and Oli L. Dors\inst{5}
    \and Ismael Garc\'{\i}a-Bernete\inst{6}
    \and Almudena Alonso-Herrero\inst{6}
    \and Anelise Audibert\inst{7,8}
    \and Enrica Bellocchi\inst{9,10}
    \and Andrew J. Bunker\inst{11}
    \and Steph Campbell\inst{12}
    \and Françoise Combes\inst{13,14}
    \and Richard I. Davies\inst{15}
    \and Tanio Díaz-Santos\inst{16,17}
    \and Fergus R. Donnan\inst{18}
    \and Federico Esposito\inst{19}
    \and Santiago Garc\'{\i}a-Burillo\inst{19}
    \and Begoña García-Lorenzo\inst{7,8}
    \and Omaira González Martín\inst{20}
    \and Houda Haidar\inst{12}
    \and Erin K. S. Hicks\inst{21,22,23}
    \and Sebastian F. Hoenig\inst{24}
    \and Masatoshi Imanishi\inst{25,26}
    \and Alvaro Labiano\inst{27}
    \and Enrique Lopez-Rodriguez\inst{28,29}
    \and Christopher Packham\inst{23}
    \and Cristina Ramos Almeida\inst{7,8}
    \and Dimitra Rigopoulou\inst{30,31}
    \and David Rosario\inst{12}
    \and Gabriel Luan Souza-Oliveira\inst{2,1}
    \and Montserrat Villar Martín\inst{1}
    \and Oscar Veenema\inst{30}
    \and Lulu Zhang\inst{23}
}

\institute{Centro de Astrobiología (CAB), CSIC-INTA, Ctra. de Ajalvir km 4, Torrejón de Ardoz, E-28850, Madrid, Spain 
 \and Departamento de F\'isica, CCNE, Universidade Federal de Santa Maria, Av. Roraima 1000, 97105-900,  Santa Maria, RS, Brazil
\and European Southern Observatory, Karl-Schwarzschild-Strasse 2, Garching bei München, Germany 
\and Instituto de Física Fundamental, CSIC, Calle Serrano 123, E-28006 Madrid, Spain 
\and Universidade do Vale do Paraíba, Av. Shishima Hifumi, 2911, Cep 12244-000, São José dos Campos, SP, Brazil
\and Centro de Astrobiología (CAB), CSIC-INTA, Camino Bajo del Castillo s/n, E-28692, Villanueva de la Cañada, Madrid, Spain  
\and Instituto de Astrofísica de Canarias, Calle Vía Láctea, s/n, E-38205, La Laguna, Tenerife, Spain 
\and Departamento de Astrofísica, Universidad de La Laguna, E-38206, La Laguna, Tenerife, Spain \ 
\and Departmento de Física de la Tierra y Astrofísica, Fac. de CC Físicas, Universidad Complutense de Madrid, E-28040 Madrid, Spain 
\and Instituto de Física de Partículas y del Cosmos IPARCOS, Fac. CC Físicas, Universidad Complutense de Madrid, E-28040 Madrid, Spain  
\and Department of Physics, University of Oxford, Denys Wilkinson Building, Keble Road, Oxford OX1 3RH, UK  
\and School of Mathematics, Statistics, and Physics, Newcastle University, Newcastle upon Tyne NE1 7RU, UK 
\and Observatoire de Paris, LUX, PSL University, Sorbonne Université, CNRS, F-75014 Paris, France 
\and Collège de France, 11 Place Marcelin Berthelot, 75231 Paris, France 
\and Max-Planck-Institut für Extraterrestrische Physik, Postfach 1312,
85741 Garching, Germany 
\and Institute of Astrophysics, Foundation for Research and TechnologyHellas, 71110 Heraklion, Greece 
\and School of Sciences, European University Cyprus, Diogenes Street, Engomi 1516, Nicosia, Cyprus 
\and Center for Astrophysics \& Space Sciences, Department of Physics, University of California San Diego, 9500 Gilman Drive, San Diego, CA 92093, USA 
\and Observatorio de Madrid, OAN-IGN, Alfonso XII, 3, E-28014 Madrid, Spain  
\and Instituto de Radioastronomía y Astrofísica (IRyA), Universidad Nacional Autónoma de México, Antigua Carretera a Pátzcuaro \#8701, Ex-Hda. San José de la Huerta, C.P. 58089 Morelia, Michoacán, Mexico 
\and Department of Physics \& Astronomy, University of Alaska Anchorage, Anchorage, AK 99508-4664, USA 
\and Department of Physics, University of Alaska, Fairbanks, Alaska 99775-5920, USA 
\and Department of Physics \& Astronomy, The University of Texas at San Antonio, One UTSA Circle, San Antonio, TX 78249, USA 
\and School of Physics \& Astronomy, University of Southampton, Southampton SO17 1BJ, UK 
\and National Astronomical Observatory of Japan, National Institutes of Natural Sciences (NINS), 2-21-1 Osawa, Mitaka, Tokyo 181-8588, Japan 
\and Department of Astronomy, School of Science, Graduate University for Advanced Studies (SOKENDAI), Mitaka, Tokyo 181-8588, Japan 
\and Telespazio UK for the European Space Agency (ESA), ESAC, Camino Bajo del Castillo s/n, 28692, Villanueva de la Cañada, Spain 
\and Department of Physics \& Astronomy, University of South Carolina, Columbia, SC 29208, USA 
\and Kavli Institute for Particle Astrophysics \& Cosmology (KIPAC), Stanford University, Stanford, CA 94305, USA 
\and Department of Physics, University of Oxford, Keble Road, Oxford OX1 3RH, UK 
\and School of Sciences, European University Cyprus, Diogenes street, Engomi, 1516 Nicosia, Cyprus 
             }


 
  \abstract
   { Active galactic nuclei (AGN), star formation (SF), and galaxy interactions can drive turbulence in the gas of the inter-stellar medium (ISM), which in turn plays a role in the SF within galaxies. The impact on molecular gas is of particular importance, as it serves as the primary fuel for SF. 
 Our goal is to investigate the origin of turbulence and the emission of molecular gas, as well as low- and intermediate-ionization gas, in the inner few kpc of both AGN hosts and star-forming galaxies (SFGs). 
 We use archival JWST MIRI/MRS observations of a sample consisting of 54 galaxies at $z<0.1$.
  Flux measurements for the  H$_2$ S(5)$\lambda6.9091\mu$m, [Ar\:{\sc ii}]$\lambda6.9853\mu$m, [Fe\:{\sc ii}]$\lambda5.3403\mu$m and [Ar\:{\sc iii}]$\lambda8.9914\mu$m emission lines along with velocity dispersion estimated by the $W_{\rm 80}$  parameter are presented.  For galaxies with coronal line emission, we include measurements for the [Mg\:{\sc v}]$\lambda5.6098\mu$m line. Line ratios are compared to photoionization and shock models to explore the origin of the gas emission. 
   AGN exhibit broader emission lines than SFGs, with the largest velocity dispersions observed in radio-strong (RS) AGN. H$_2$ gas is less turbulent compared to ionized gas, while coronal gas presents higher velocity dispersions. The $W_{\rm 80}$ values for the ionized gas exhibits a decrease from the nucleus out to radii of approximately 0.5--1 kpc, followed by an outward increase up to 2--3 kpc. In contrast, the H$_2$ line widths generally display increasing profiles with distance from the center. 
 Correlations between the $W_{\rm 80}$ parameter and line ratios such as H$_2$ S(5)/[Ar\,{\sc ii}] and [Fe\,{\sc ii}]/[Ar\,{\sc ii}] indicate that the most turbulent gas is associated with shocks, enhancing H$_2$ and [Fe\,{\sc ii}] emissions.   
   Based on the observed line ratios and velocity dispersions, the [Fe\:{\sc ii}] emission is consistent with predictions of fast shock models, while the H$_2$ emission is likely associated with molecules formed in the post-shock region.  We speculate that these shocked gas regions are produced by AGN outflows and jet-cloud interactions in AGN-dominated sources, while in SFGs, they may be created by stellar winds and mergers.  This shock-induced gas heating may be an important mechanism of AGN (or stellar) feedback, preventing the gas from cooling and forming new stars.
}

   \maketitle
%

\section{Introduction}
The interplay between Active Galactic Nuclei (AGN) or stellar-driven galactic winds and star formation (SF) in galaxies is a crucial aspect of galaxy evolution. Both AGN and stellar winds can significantly suppress or regulate SF by injecting large amounts of energy into the interstellar medium \citep[ISM; e.g. ][]{silk98,DiMatteo05,Hopkins12,dalla-vecchia12,Heckman14,Fierlinger16,2017Harrison_nature,Harrison24,Veilleux20,Silk24}. The strong radiation field from AGN or nuclear starburst, nuclear winds, and jets can all generate turbulence in the gas and induce shocks, which, in turn, can disrupt the dense gas necessary for SF. These dynamic processes hinder or even prevent SF (i.e., negative feedback), as they inhibit the gas from condensing and cooling effectively. However, these mechanisms can also promote SF (i.e., positive feedback). For instance, fast outflows may enhance SF by compressing molecular clouds under pressure or by fostering SF within the outflowing material itself \citep{Silk13,Zubovas14,Cresci15,Maiolino17,Gallagher19,Garcia-Bernete21,Bessiere22,Hermosa24}. Consequently, the balance between these outflows and the gas available for SF can have a profound impact on the star-forming potential of galaxies.  A central problem is therefore to understand the origin of the gas turbulence observed in the central regions of galaxies and to establish the role of shocks in driving it. Determining whether turbulence is primarily induced by AGN outflows, radio jets, or stellar feedback, and how it regulates the physical state of the ISM, is essential to assess its impact on galaxy evolution. Addressing this issue is crucial for disentangling the mechanisms of feedback that suppress or promote star formation in different environments.

Shock-induced turbulence, triggered by AGN activity or stellar winds, can excite a variety of transitions in molecular and ionized gas, enhancing its emission across multiple wavelengths \citep[e.g.][]{Dopita95,Contini01,allen08,dors21_suma,rogemar_shocks21,venturi21,Appleton17,Appleton23,Audibert23,Audibert25,Astor25,Ardila25}. Understanding gas emission mechanisms and the resulting turbulence in the central regions of galaxies is crucial for advancing our knowledge of how feedback processes, driven by both stellar activity and AGN, contribute to the evolution of galaxies.

Observations of the ionized gas kinematics using integral field spectroscopy (IFS) show that AGN exhibit higher [O\:{\sc iii}]$\lambda5007$ emission line widths compared to star-forming galaxies (SFGs), extending up to distances greater than one effective radius ($R_{\rm e}$) from the nucleus, and that high-luminosity AGN, with $L_{\rm [O~III]} > 2\times10^{40}\ {\rm erg\ s^{-1}}$, display a sharp increase in line widths within $0.4\,R_{\rm e}$, which is associated with AGN-driven outflows \citep{wylezalek20,deconto-machado22,Gatto24}.  Evidence of increased turbulence is detected in luminous and ultra-luminous infrared galaxies (U/LIRGs) using optical emission lines \citep[e.g.][]{Bellocchi13,Arribas14,Perna22}. \citet{Alban24} found that radio-selected AGN exhibit broader [O\:{\sc iii}]$\lambda$5007 lines compared to those selected by other techniques, such as optical or infrared (IR) diagnostics. The authors conclude that this difference arises because radio-selected AGN represent a population in which AGN-driven kinematic perturbations have been active for longer durations, consistent with the radio emission being driven by shocks from outflows.  Studies of the ionized gas in the inner kpc of AGN hosts show that broadened or complex emission line profiles are generally associated with outflows or turbulence induced by the interaction of radio jets or outflows with the ISM \citep[e.g.][]{Mullaney13,fischer18,freitas18,forster_schreiber19,avery21,kakkad22,Zhang24,Davies24,Speranza24,Esposito24,Garcia-bernete24,Hermosa24}. In some cases, this interaction results in additional emission from gas excited by shocks \citep[e.g.][]{rogemar_shocks21,venturi21}.

Vibrational and ro-vibrational H$_2$ emission in the central few kpc of nearby AGN hosts and ULIRGs can be strongly enhanced by shocks from outflows or jet--cloud interactions \citep[e.g.,][]{Ogle10,ogle25,hill14,colina15,Kristensen23,rogemar_21_exc,rogemar21_survey,rogemar25_jwst,henrique24,Villar-Martin23,Leftley24,Bohn24,Dasyra24}. These shocks can heat the gas, promoting the excitation of H$_2$ molecules and leading to the emission of these diagnostic lines. A comparison of the emission lines from hot molecular gas and ionized gas, using near-IR lines, shows that the former are typically narrower than the latter, but both are observed in kinematically disturbed regions (KDRs) around AGN and in ULIRGs \citep[e.g.][]{emonts17,RamosAlmeida17,Ramos-Almeida19,RamosAlmeida25,bianchin22,rogemar23_extended_kin,Zanchettin25}.

With the use of  the James Webb Space Telescope  \citep[JWST; ][]{Gardner23}  Mid-Infrared Instrument medium-resolution spectrometer \citep[MIRI/MRS; ][]{Wright15,Wright23}, it is possible to map the kinematics of multiple gas phases in the central regions of galaxies, including areas that are heavily obscured at optical wavelengths, as the extinction in the mid-IR is up to $\sim$40 times lower than in the optical \citep{Gordon23}.  This allows unprecedented access to the deeply embedded gas emission structures and processes in dusty galactic nuclei, such as AGN-driven winds, turbulence in the ISM produced by outflows and jets, and circumnuclear SF. Here, we use archival JWST MIRI/MRS observations of a large sample of AGN hosts and ULIRGs at redshifts  $z<0.1$ to compare the kinematics across warm molecular gas and low- to moderate-ionization phases, and investigate the origin of the gas emission and turbulence in these galaxies. 
 We focus on the H$_2\,0-0$ S(5)$\lambda6.9091\mu$m transition and a set of low- and medium-ionization fine-structure lines, including [Ar\,{\sc ii}]$\lambda6.99\,\mu$m, [Fe\,{\sc ii}]$\lambda5.34\,\mu$m, and [Ar\,{\sc iii}]$\lambda8.99\,\mu$m. These lines are among the brightest in the mid-IR spectra, and they often trace spatially extended emission in nearby AGN and SFGs. Their excitation can arise from different mechanisms, such as photoionization by massive stars or AGN, as well as shocks produced by stellar and AGN-driven outflows and jets. The analysis of both the line intensities and their kinematics therefore provides powerful diagnostics of the physical conditions and dominant excitation processes in the nuclear regions of galaxies.

 This paper is organized as follows. Section 2 describes the sample, and Section 3 presents the data and measurements. Our main results are presented in Section 4, followed by a discussion on the origin of the gas emission and turbulence in Section 5. Finally, Section 6 summarizes our key conclusions.

\begin{table*}
    \caption[]{Properties of the sample.}
    \label{tab:lums}
\centering
\renewcommand{\arraystretch}{0.9}  
\fontsize{8.5pt}{9.8pt}\selectfont  
\begin{tabular}{|l|c|c|c|c|c|c|l|}
\hline
(1) & (2) & (3) & (4) & (5) & (6) & (7) & (8) \\
 Object & $D$ & log $L_{\rm BAT}$ & log $P_{\rm 1.4 GHz}$  & log $P_{\rm 4.8 GHz}$ & log $L_{\rm H}$  & log $L_{\rm IR}$  & Subsamples \\
  & [Mpc] & [erg/s] & [W/Hz]  & [W/Hz] & [erg/s] & [erg/s] & \\
\hline
Arp220 & 78.8 & -- & 23.36 & 23.20 & 43.19 & 45.68 & SF\\
Centaurus A & 3.1$^a$ & 42.19 & 21.71$^a$ & 21.83$^a$ & 41.19 & 43.25 & BAT; [Mg\:V] \\
Cygnus A & 240.9 & 44.98 & 26.74$^b$ & 27.15$^b$ & 43.58 & 45.79 & BAT; RS; [Mg\:V]; IR \\
ESO137-G034 & 31.0$^b$ & 42.43 & -- & -- & 42.54 & 44.05 & BAT; [Mg\:V] \\
ESO420-G13 & 51.0 & -- & 22.30 & -- & 43.06 & 44.95 &  [Mg\:V]; IR  \\
IC5063 & 48.6 & 43.28 & 23.73$^c$ & 23.17$^c$ & 43.06 & 45.13 & BAT; RS; [Mg\:V]; IR  \\
IIZw96 & 154.7 & -- & 23.08 & 22.26$^d$ & 42.72 & 45.82 &  IR  \\
IRAS05189-2524 & 188.8 & 43.60 & 23.07 & -- & 44.13 & 46.13  & BAT; IR  \\
IRAS07251-0248 & 375.2 & -- & 23.27 & -- & 43.73 & 45.91 & IR  \\
IRAS09022-3615 & 255.6 & -- & 23.80 & -- & 43.77 & 45.92 & RS; IR  \\
IRAS09111-1007 & 232.0 & -- & 22.26 & -- & 43.81 & 45.38 & SF\\
IRAS10565+2448 & 184.7 & -- & 23.35 & -- & 43.91 & 45.68 & SF \\
IRAS13120-5453 & 131.8 & -- & 23.49$^c$ & 23.12$^c$ & 43.96 & 45.91 & SF \\
IRAS14348-1447 & 352.8 & -- & 23.69 & -- & 43.74 & 45.00 &  SF \\
IRAS15250+3609 & 236.6 & -- & 22.96 & -- & 43.56 & 45.89 & IR  \\
IRAS19297-0406 & 367.4 & -- & 23.64 & -- & 43.89 & 45.85 & SF\\
IRAS19542+1110 & 278.4 & -- & 23.25 & -- & 43.97 & 45.72 & SF \\
IRAS20551-4250 & 184.3 & -- & 23.08$^e$ & -- & 43.58 & 45.85 & IR  \\
IRAS22491-1808 & 333.2 & -- & 22.86 & -- & 43.61 & 45.78 & SF \\
IRAS23128-5919 & 191.1 & -- & -- & -- & 43.31 & 45.85  & IR \\
IRASF01364-1042 & 206.7 & -- & 22.87 & 22.76 & 43.26 & 45.20 & IR  \\
IRASF08572+3915NW & 249.4 & -- & 22.48 & -- & 43.09 & 46.07 &  IR  \\
IRASF14378-3651 & 289.9 & -- & 23.50 & -- & 43.71 & 45.68 & IR  \\
IRASF23365+3604 & 276.3 & -- & 23.37 & -- & 43.78 & 45.81 & IR \\
M81 & 3.3$^c$ & 40.42 & 20.05 & 20.11 & 41.80 & 42.32  & BAT\\
M87 & 15.8$^d$ & -- & 24.64 & 24.24 & 42.23 & 42.79 & RS\\
M94 & 4.2$^e$ & -- & 19.41 & 20.35 & 42.09 & 42.84 & SF\\
M104 & 8.0$^f$ & -- & 20.86 & 21.08$^f$ & 42.43 & 42.37 & SF \\
MCG-05-23-016 & 36.6 & 43.52 & 21.36 & -- & 42.97 & 44.70  & BAT; [Mg\:V]; IR  \\
Mrk231 & 180.7 & -- & 24.06 & 24.19 & 44.81 & 46.46 & RS; IR \\
Mrk273 & 160.0 & 43.18 & 23.63 & 23.33 & 43.73 & 45.81  & BAT; [Mg\:V]; IR  \\
NGC0253 & 3.5$^e$ & -- & 21.59 & 21.55$^h$ & 41.38 & 44.37 & SF \\
NGC0424 & 50.4 & 42.81 & 21.84 & -- & 43.23 & 44.86 & BAT; IR  \\
NGC1052 & 19.2$^g$ & 42.14 & 22.60 & 23.14$^f$ & 42.75 & 43.63  & BAT \\
NGC1068 & 10.1$^h$ & 41.66 & 22.77 & 22.29$^g$ & 42.90 & 45.23   & BAT; [Mg\:V]; IR \\
NGC1365 & 18.3$^i$ & 42.40 & 22.18 & 22.08$^h$ & 42.00 & 44.90   & BAT; [Mg\:V]\\
NGC1566 & 6.6$^g$ & 41.01 & 20.93$^c$ & 20.72$^i$ & 41.62 & 42.71    & BAT; [Mg\:V]\\
NGC1808 & 9.5$^j$ & -- & 21.76 & 21.41$^i$ & 42.21 & 44.47       & SF\\
NGC3081 & 23.8$^k$ & 42.74 & 20.56 & -- & 42.30 & 44.10  & BAT; [Mg\:V]\\
NGC3256N & 40.1 & -- & -- & -- & 43.06 & 45.60  & SF \\
NGC3256S & 40.1 & -- & 23.07$^c$ & 22.78$^i$ & 43.17 & -- & IR \\
NGC4258 & 6.8$^l$ & 41.11 & 21.44 & 21.23$^b$ & 41.80 & 42.76  & BAT\\
NGC4395 & 4.4$^m$ & 40.81 & 19.75 & 18.07$^i$ & 39.78 & 41.61  & BAT; [Mg\:V]\\
NGC5506 & 23.8$^n$ & 43.21 & 22.36 & 22.09$^k$ & 42.92 & 44.57  & BAT; [Mg\:V]; IR \\
NGC5728 & 39.0$^o$ & 43.18 & 22.10 & -- & 42.86 & 44.36  & BAT; [Mg\:V]\\
NGC6240 & 104.2 & 43.94 & 23.73 & 23.32 & 43.89 & 45.66  & BAT, RS; [Mg\:V]\\
NGC6552 & 113.7 & 43.50 & 22.68 & 22.70$^d$ & 43.48 & 45.25  & BAT; [Mg\:V]; IR \\
NGC7172 & 37.4 & 43.42 & 21.79 & -- & 42.88 & 44.20  & BAT; [Mg\:V]; IR \\
NGC7319 & 96.5 & 43.60 & 22.76 & -- & 43.11 & 44.59  & BAT; [Mg\:V]; IR \\
NGC7469 & 69.7 & 43.61 & 23.01 & 22.60 & 43.75 & 45.56  & BAT; [Mg\:V]\\
NGC7582 & 23.2 & 42.72 & 22.22$^e$ & 21.68$^k$ & 43.02 & 44.77  & BAT; [Mg\:V]; IR \\
UGC05101 & 168.7 & 43.37 & 23.75 & 23.40 & 43.88 & 45.54  & BAT, RS; IR \\
VV114 & 86.0 & -- & 23.33 & 22.92$^f$ & 43.20 & 45.58  & IR \\
VV340a & 144.3 & -- & 23.34 & 22.81 & 43.87 & 45.05  & [Mg\:V]\\
\hline
\end{tabular}
\vspace{-0.2cm}
\tablefoot{ (1) Galaxy name;
(2) Distances were taken from the literature when available; otherwise, they were estimated from the redshift, assuming a cosmology with $h = 0.7$, $\Omega_{m} = 0.3$, and $\Omega_{\Lambda} = 0.7$. Superscript letters indicate the corresponding references: 
a: \citet{Majaess08}, 
b: \citet{Willick97}, 
c: \citet{Gerke11}, 
d: \citet{Oldham16}, 
e: \citet{Radburn-Smith11}, 
f: \citet{Spitler06}, 
g: \citet{Tully13}, 
h: \citet{Nasonova11}, 
i: \citet{Riess16}, 
j: \citet{tully16}, 
k: \citet{Bottinelli84}, 
l: \citet{Hoffmann15}, 
m: \citet{Sabbi18}, 
n: \citet{Karachentsev14}, 
o: \citet{Rest14}; 
(3) Hard X-ray (14-195 keV) luminosity from \citet{BAT105}; 
(4) 1.4 GHz power taken from the NRAO VLA Sky Survey \citep[NVSS; ][]{Condon98}, and supplemented with additional data from the literature (indicated by superscripts): 
a: \citet{Tingay03}, 
b: \citet{Steenbrugge10},  
c: \citet{Allison14}, 
d: \citet{Brown17}, 
e: \citet{Condon96}; 
 (5) 4.8 GHz radio power taken from the Green Bank 4.85 GHz survey \citep{Gregory96}, with additional data from the literature (indicated by superscripts):  
a: \citet{Tingay03}, 
b: \citet{Becker91}, 
c: \citet{Wright94}, 
d: \citet{Gregory91},  
e: \citet{Baan06}, 
f: \citet{Griffith94},  
g: \citet{Sajina11}, 
h: \citet{Stil09}, 
i: \citet{Wright94},  
j: \citet{Nagar05}, 
k: \citet{Orienti10}.
(6) H-band luminosity from The Two Micron All Sky Survey \citep[2MASS;][]{Skrutskie06}; 
(7) IR luminosity obtained from the WISE W4 luminosity \citep{Wright10} using the relation from 
\citet{Cluver17}. The northern nucleus of NGC~3256 is an order of magnitude brighter in IR than the southern nucleus \citep{Bohn24}; thus, we quote $L_{\rm IR}$ only for the northern nucleus. High-resolution IR data reveal an AGN in the southern nucleus \citep{Ohyama15}, which is included in our IR AGN sample. For IRAS 13120-5453, $L_{\rm IR}$ was estimated from the 25$\mu$m IRAS flux, as WISE magnitudes are unavailable; 
(8) The subsamples to which each object belongs.
}
\renewcommand{\arraystretch}{1}  
\end{table*}

\begin{table*}[!ht]
    \caption[]{Mean properties for each subsample. Uncertainties are the standard deviation of each parameter.}
    \label{tab:mean}
\centering
\small
\begin{tabular}{|c|c|c|c|c|c|}
\hline
Subsample & BAT AGN & RS AGN & [Mg\:{\sc v}] AGN & IR AGN & SF \\
\hline
$\#$ of galaxies &  25 & 7& 21 & 27 & 13 \\
$\langle \log D/{\rm Mpc}\rangle$ & $1.50\pm0.56$ &  $2.03\pm0.44$ &  $1.55\pm0.51$ & $2.02\pm0.41$ &    $1.80\pm0.78$ \\
$\langle \log L_{\rm H}/{\rm erg\:s^{-1}}\rangle$& $42.79\pm0.98$ & $43.60\pm0.80$ & $42.81\pm0.97$ & $43.39\pm0.47$ & $43.17\pm0.87$\\
$\langle \log L_{\rm IR}/{\rm erg\:s^{-1}}\rangle$& $44.46\pm1.18$ & $45.33\pm1.19$ & $44.58\pm1.04$ & $45.45\pm0.57$ & $44.97\pm1.17$\\
$\langle \log P_{\rm 1.4}/{\rm W\:Hz^{-1}}\rangle$& $22.23 \pm 1.65$ & $24.35 \pm 1.11$ & $22.42 \pm 1.60$&$23.07 \pm 1.00$ & $22.51 \pm 1.29$ \\
\hline
\end{tabular}
\end{table*}

\section{The sample}

 Our goal is to investigate the origin of the emission and the dynamics of warm molecular gas as well as low- and medium-ionization ionized gas. 
 In local galaxies ($z\lesssim0.1$), the aforementioned emission lines are observed in Channels 1 and 2 of the MIRI/MRS, which provide angular resolutions of approximately 0.35--0.40 arcsec (FWHM; \citealt{Law23}).  Additionally, we used the [Mg\,{\sc v}]$\lambda5.6098\,\mu$m emission line to identify objects with high-ionization gas, and H$_2$ S(3)$\lambda9.6649\,\mu$m to investigate the origin of the molecular gas emission.  The ionization potentials are 7.9 eV for Fe\:{\sc ii} \citep{Nave13}, 15.8 eV for Ar\:{\sc ii} \citep{Sansonetti05}, 27.6 eV for Ar\:{\sc iii} \citep{Kaufman96}, and 109 eV for Mg\:{\sc v} \citep{Biemont99}, as listed in the National Institute of Standards and Technology (NIST) Atomic Spectra Database Ionization Energies Data \citep{Kramida14}.  

 The sample was defined as follows: we queried the Mikulski Archive for Space 
Telescopes (MAST) Portal for JWST observations of galaxies with $z<0.1$ 
obtained using the MIRI instrument in the MRS observing mode. The redshift threshold was selected to allow the study of gas emission and kinematics on scales of hundreds of parsecs, enabling meaningful comparisons across objects with similar spatial resolutions.  Our search was limited to projects with publicly available data that provide complete spectral coverage with MIRI/MRS, including all three sub-bands (Short, Medium, and Long) and extended emission in the relevant emission lines, resulting in a sample of 54 galaxies. Table~\ref{tab:lums} summarizes the properties of the sample, including the distance ($D$), hard X-ray (14–195 keV) luminosity ($L_{\rm BAT}$), radio powers at 1.4 GHz and 4.8 GHz ($P_{\rm 1.4\,GHz}$ and $P_{\rm 4.8\,GHz}$), H-band luminosity ($L_{\rm H}$), and infrared luminosity ($L_{\rm IR}$) of each galaxy. 

In our sample, 25 objects (46\%) are detected in the Swift BAT survey, while 51 objects (94\%) have 1.4 GHz data, and 30 objects (56\%) have measurements at 4.8 GHz (see notes in Table~\ref{tab:lums} for references).  Table~\ref{tab:sample} presents the list of galaxies and the corresponding details of the observational proposals. Strong correlations are found between radio power and X-ray luminosity for the objects detected in both bands. Specifically, the 1.4~GHz and 4.8~GHz radio powers show Pearson correlation coefficients of 0.79 and 0.85, respectively, with the X-ray luminosity. Additionally, for the objects detected at both radio frequencies, the 4.8~GHz and 1.4~GHz radio powers are tightly correlated, closely following a 1:1 relation, with a Pearson correlation coefficient of 0.96.

 In this work, we are not focused on discussing individual objects, many of which exhibit a rich abundance of physical properties.  Instead, our goal is to investigate the general properties of the sample, specifically the physical characteristics of warm molecular gas and low- to moderate-ionization gas. Consequently, we do not present individual maps for all galaxies. We present our results by dividing the sample into five subsamples, where a single galaxy may belong to more than one group, defined as follows:

\begin{itemize}

    \item  BAT AGN: This subsample of X-ray selected AGN is composed of 25 galaxies. It is obtained by cross-matching the sample with the 105-month catalog of hard X-ray sources (14–195 keV) from the Swift Burst Alert Telescope (BAT) survey \citep{BAT105}, including all X-ray detected sources. The hard X-ray emission offers a direct measurement of AGN activity, as it predominantly captures the intrinsic emission from the AGN rather than scattered or reprocessed emissions. Additionally, it is significantly less affected by line-of-sight obscuration compared to optical wavelengths or softer X-ray bands. 
    
    \item RS AGN: This subsample of radio-strong (RS) AGN is composed of 7 galaxies. SFGs typically exhibit 1.4 GHz radio luminosities below $P_{\rm 1.4 GHz}=10^{23}\:{\rm W\:Hz^{-1}}$, although some can reach up to $\sim10^{24}\:{\rm W\:Hz^{-1}}$ \citep{Condon02,Jose24}. AGN, on average, exhibit radio luminosities higher than those expected purely from SFGs \citep[e.g.][]{Condon02}, due to additional emission from jets \citep{Padovani17} and/or shocks generated by outflows \citep{Zakamska16}. The standard definition of radio-loud AGN identifies them as those with a ratio of 5 GHz radio luminosity to B-band luminosity of $L_{\text{5GHz}}/{L_{\rm B}} \gtrsim 10$ \citep[e.g.][]{Kellermann89}. However, several studies use the 1.4 GHz radio power to distinguish between radio-loud and radio-quiet AGN, often defining thresholds at $P_{\rm 1.4 GHz}=10^{23}\:{\rm W\:Hz^{-1}}$ \citep[e.g.][]{Best05} or $P_{\rm 1.4 GHz}=10^{24}\:{\rm W\:Hz^{-1}}$ \citep[e.g.][]{Tadhunter16}.  Since most galaxies in our sample are detected at 1.4 GHz, we use their observed radio powers to define a subsample of RS AGN. Only three objects in our sample have $\log P_{\rm 1.4\,GHz}/[\rm W\,Hz^{-1}] > 24$.  On the other hand, our sample includes ULIRGs, for which the commonly used  threshold of $\log P_{\rm 1.4\,GHz}/[\rm W\,Hz^{-1}] > 23$ may be insufficient to ensure the presence of an AGN. Therefore, we adopt an intermediate value of $\log P_{\rm 1.4\,GHz}/[\rm W\,Hz^{-1}] > 23.7$, which lies between the thresholds commonly used in the literature.  This sample includes four ULIRGs (i.e. IRAS 09022$-$3615, Mrk~231, NGC~6240, and UGC~05101) all of which are classified as AGN  using the other methods adopted in this work.  In these objects, a significant fraction of the radio emission may still be attributed to SF.  Among all the RS AGN, the only galaxy not included in any other AGN sample is the radio galaxy M\:87.  In addition, we note that only two galaxies in this sample exhibit a clear radio excess: Cygnus\,A and M\,87. This is evidenced by $q_{\rm 23} < 0$, where $q_{\rm 23} = \log(S_{\rm 23\mu m}/S_{\rm 1.4GHz})$, with $S_{\rm 23\mu m}$ and $S_{\rm 1.4GHz}$ representing the flux densities at 23 $\mu$m and 1.4 GHz, respectively \citep{Radcliffe21}. Therefore, interpretations regarding the role of AGN in this subsample should be approached with caution.
    
   \item {[Mg\:{\sc v}] AGN:} This subsample is composed of 21 galaxies with coronal line emission.  Coronal lines are emitted by highly ionized gas with ionization potentials IP$\gtrsim$100 eV. They serve as reliable indicators of AGN activity, generated either by photoionization from the AGN's intense radiation field or by shocks associated with jets and outflows \citep[e.g.][]{Ardila25,RamosAlmeida25}. We use the [Mg\:{\sc v}]$\lambda5.6098\mu$m emission line to select galaxies for this subsample, using the MIRI MRS datacubes. We show the [Mg\:{\sc v}] flux maps and line profiles for each galaxy in Fig.~\ref{fig:mgv}. 

    \item  IR AGN:  This subsample is composed of 27 IR selected galaxies. AGN exhibit an excess of IR emission, which arises from the dusty torus heated by the radiation from the central engine \citep{Antonucci93,Netzer15}. We select the IR AGN sample based on magnitudes from the Wide-field Infrared Survey Explorer \citep[WISE;][]{Wright10}.  We adopt the definition of \citet{Stern12}, considering AGN as objects  with a color criterion of $W1-W2 \geq 0.8$. This criterion efficiently identifies AGN even in galaxies with significant host galaxy contamination and is particularly sensitive to dust-obscured AGN populations \citep{Assef10,Stern12}.

    \item SF: The subsample comprises 13 SFGs that are not classified as AGN by any of the methods described above, and is predominantly composed of U/LIRGs (7 ULIRGs and 2 LIRGs). Although some of these galaxies show evidence of buried or obscured AGN, their gas excitation is predominantly driven by SF  \citep{cicone14,su23,Garcia-Bernete25}. 

\end{itemize}

 Table~\ref{tab:mean} presents the mean properties of each subsample. The H-band luminosity can be used as a reliable proxy for stellar mass \citep{davies15}, allowing us to compare the stellar mass distributions across different subsamples. 
The  sample includes objects with $L_{\rm H}$ in the range 10$^6$-10$^{\rm 11}$\:L$_\odot$. 
The mean H-band luminosities for all subsamples are similar, with the BAT AGN and [Mr\,{\sc v}] AGN subsamples presenting slightly smaller values than the other samples. In addition, we performed two-sample Kolmogorov–Smirnov (KS) tests to assess whether the H-band luminosity distributions of the AGN and SF samples differ significantly. The p-values indicate that the AGN and SF samples are drawn from the same parent population.

\begin{figure}[ht]
   \centering
   \includegraphics[width=0.45\textwidth]{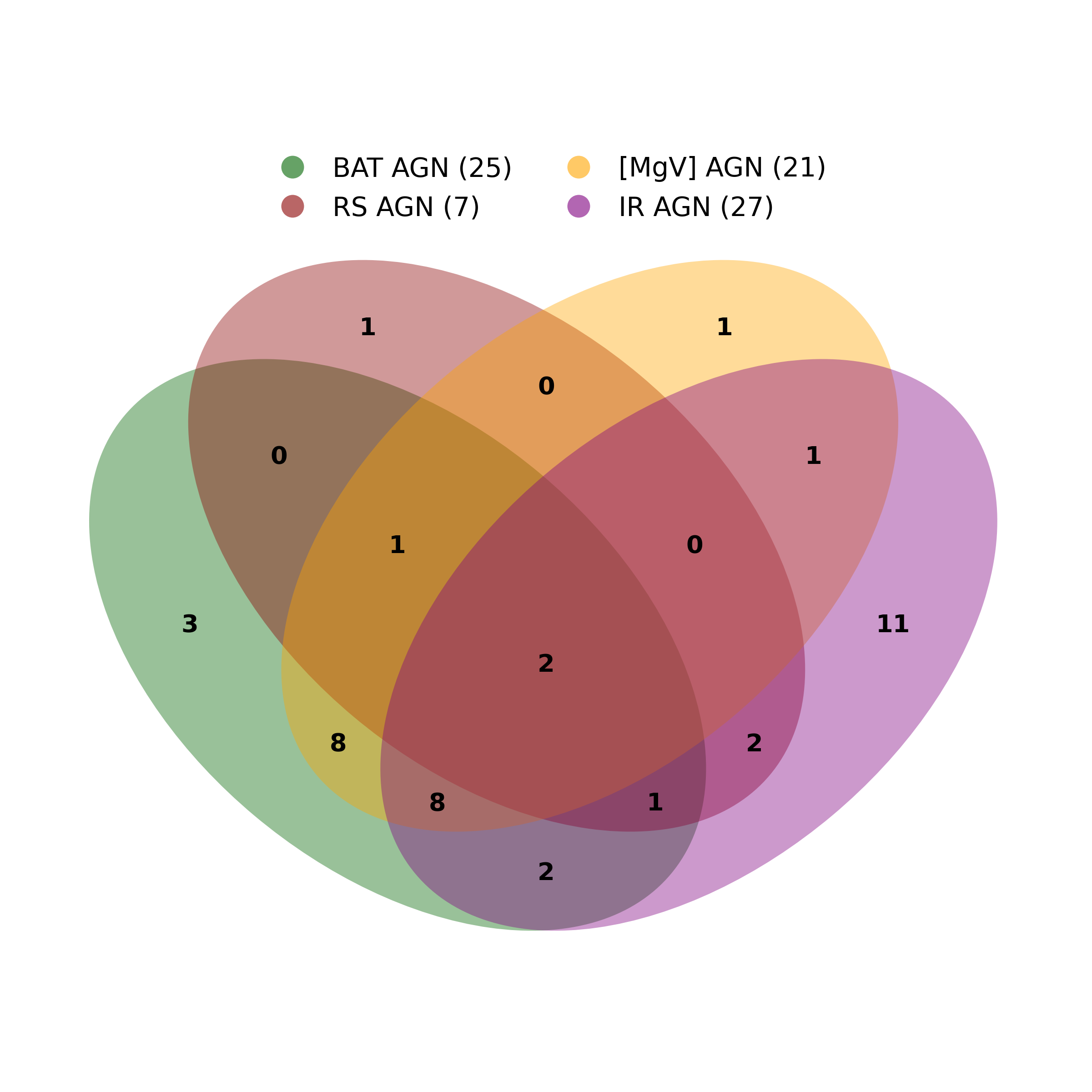}
   \vspace{-1cm}
         \caption{Venn diagram illustrating the overlap among the different AGN subsamples.}              
         \label{fig:venn}
   \end{figure}

 The four AGN subsamples were defined using tracers of different physical processes. 
The BAT subsample directly probes the emission of the hot corona located above the AGN accretion disk, while the RS subsample is associated with the mechanical energy released in the form of jets. 
The [\ion{Mg}{v}] subsample is related to the local gas physics, as it primarily traces the ionization parameter rather than directly probing the AGN emission, similar to optical diagnostic diagrams \citep{bpt_1981,Negus23}. 
Coronal line emission, however, can also be associated with shocked gas regions in the inner $\sim$1\,kpc of AGN hosts \citep{Ardila25,Ardila25b}. 
Finally, the IR subsample traces the physics of the hot dust heated by AGN radiation. 
Since these physical processes are interconnected and can occur simultaneously in an AGN, many of our objects appear in more than one subsample. Fig.~\ref{fig:venn} shows the Venn diagram illustrating the overlap among the different AGN subsamples.

   \begin{figure}[ht]
   \centering
   \includegraphics[width=0.45\textwidth]{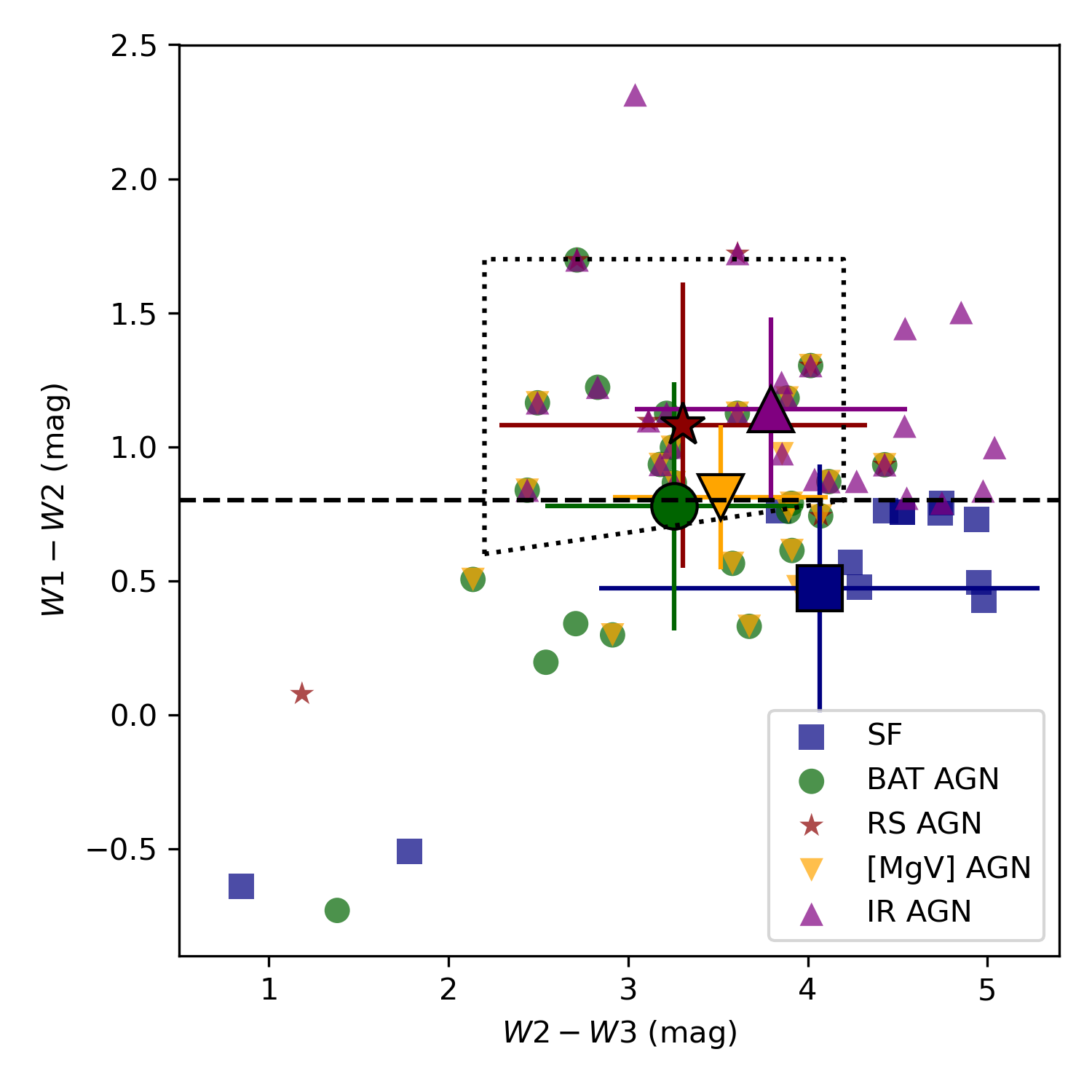}
         \caption{WISE color–color diagram showing the five subsamples, as indicated by the symbols. The uncertainties are comparable to the symbol sizes. The large symbols with error bars correspond to the mean values for each subsample, with the error bars representing the standard deviation. The dashed line represents the threshold $W1-W2 = 0.8$, while the dotted polygon marks the region typically occupied by Seyfert galaxies \citep{Jarrett11}.}      
         \label{fig:wise}
   \end{figure}


 In Fig.\ref{fig:wise} we present the WISE color-color diagram for our subsamples. The mean $W1-W2$ and $W2-W3$ values for all AGN subsamples fall within the region typically occupied by Seyfert galaxies, as indicated by the dotted polygon. The IR AGN subsample includes representatives of dustier objects, exhibiting larger colors along both axes of the diagram; some BAT and [\ion{Mg}{v}] AGN show lower $W1-W2$ values than typical Seyfert galaxies, consistent with low-luminosity AGN; and objects in the SF sample are located in a region consistent with Starburst galaxies; all three associations are consistent with the locations of these classes among WISE objects \citep{Wright10}.

   \begin{figure*}
   \centering
   \includegraphics[width=0.99\textwidth]{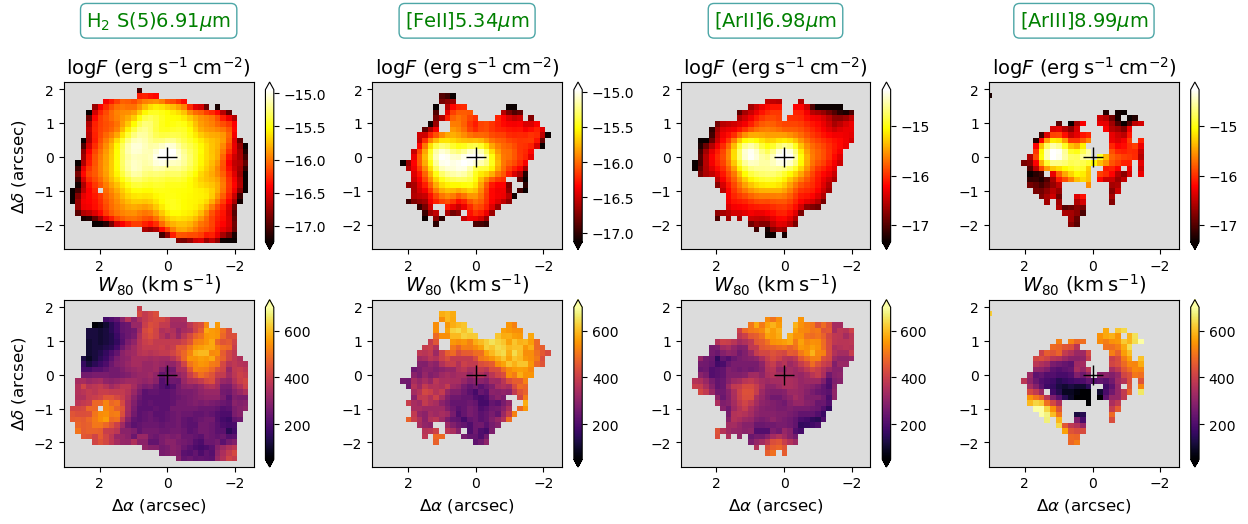}
         \caption{Examples of flux (top panels) and $W_{\rm 80}$ (bottom panels) maps for Arp 220. From left to right, the H$_2$ S(5)$\lambda6.9091\mu$m, [Ar\:{\sc ii}]$\lambda6.9853\mu$m, [Fe\:{\sc ii}]$\lambda5.3403\mu$m, and [Ar\:{\sc iii}]$\lambda8.9914\mu$m  are shown. The central crosses identify the location of the peak of the continuum, corresponting to the western nucleus, used as reference to calculate radial distances from the galaxy nucleus. The gray regions correspond to locations where the corresponding emission line is not detected with $snr > 5$ and regions not covered by the MRS FoV.}              
         \label{fig:maps}
   \end{figure*}

\section{Data Reduction and Measurements} 

We use archival mid-IR JWST spectroscopic data obtained with the MIRI/MRS instrument \citep{2015PASP..127..646W,Labiano21,2023A&A...675A.111A} of a sample of nearby galaxies. The observations were carried out using different observational strategies, according to the scientific objectives of the approved proposals. We downloaded the processed data from MAST archive, using the following filters: \texttt{obs\_collection='JWST', intentType='science', dataRights='PUBLIC', instrument\_name='MIRI/IFU', calib\_level=3, proposal\_id=pid, obs\_id='jw*\{pid\}-c*'}, where \texttt{pid} refers to the proposal IDs listed in the table, identified from the list of approved projects up to cycle 3. These data were processed using version 1.16.1 of the JWST Science Calibration Pipeline \citep{bushouse_2024}, employing the reference file \texttt{jwst\_1303.pmap}.

The MIRI/MRS field of view (FoV) for channel 1 is 3.2$\times$3.7 arcsec$^2$, while channel 2 has a FoV of 4.0$\times$4.8 arcsec$^2$. 
 Before performing the emission line flux and kinematic measurements, we rebinned the channel 1 data cubes to a spaxel size of 0.17 arcsec and convolved them with a Gaussian function with FWHM = 0.40 arcsec, in order to match the angular sampling and spatial resolution of the channel 2 cubes. Subsequently, each spaxel in the cube was replaced with the average of its nearest neighbors within a 3$\times$3 spaxel box. This process minimizes residual instrumental effects, particularly the continuum wiggles caused by the undersampling of the point spread function  \citep[PSF;][]{Law23}.

We performed the flux measurements and calculated the velocity dispersion, parametrized by the $W_{\rm 80}$ parameter, which is defined as the width encompassing 80 per cent of the total line flux.  These measurements were carried out for the following emission lines: H$_2$ S(5)$\lambda6.9091\mu$m, H$_2$ S(3)$\lambda9.6649\mu$m, [Ar\:{\sc ii}]$\lambda6.9853\mu$m, and [Ar\:{\sc iii}]$\lambda8.9914\mu$m, [Fe\:{\sc ii}]$\lambda5.3403\mu$m and [Mg\:{\sc v}]$\lambda5.6098\mu$m.  
For each emission line, we subtracted the continuum contribution by fitting a linear function to regions adjacent to the line. The line profile was then integrated within a 3000 km\:s$^{-1}$ window centered on the emission line. We only considered spaxels where the emission line was detected with a signal-to-noise ratio $snr > 5$, determined as the ratio between the line profile amplitude and the standard deviation of the adjacent continuum within a 1000 km\:s$^{-1}$ window. Fig.~\ref{fig:maps} presents examples of flux and $W_{\rm 80}$ measurements for the emission lines H$_2$ S(5)$\lambda6.9091\mu$m, [Ar\:{\sc ii}]$\lambda6.9853\mu$m, [Fe\:{\sc ii}]$\lambda5.3403\mu$m, and [Ar\:{\sc iii}]$\lambda8.9914\mu$m in the galaxy Arp 220. The gray regions indicate areas where the corresponding emission line is either not detected with $snr > 5$ or lies outside the FoV of the MIRI instrument. The measured $W_{\rm 80}$ values were corrected for instrumental broadening using the resolving power described by $R = 4603 - 128 \times \lambda [\mu{\rm m}]$ \citep{Jones23}, along with the relation $W_{\rm 80} = 1.09 \times {\rm FWHM}$ for a Gaussian profile.

 \section{Results}

   \begin{figure*}[ht]
   \centering
   \includegraphics[width=0.99\textwidth]{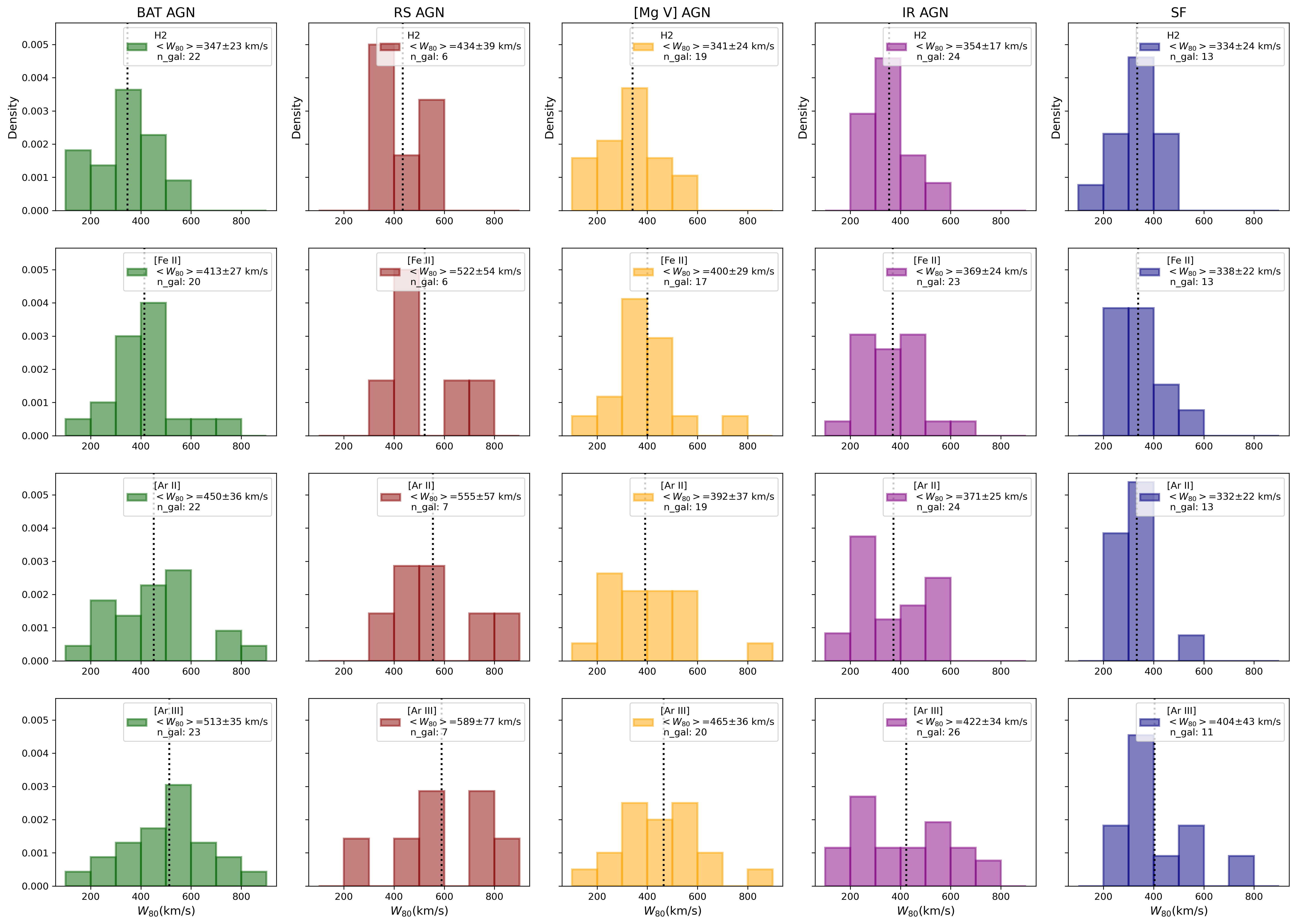}
         \caption{Nuclear $W_{\rm 80}$ values for the H$_2$ S(5), [Fe\,\textsc{ii}], [Ar\,\textsc{ii}], and [Ar\,\textsc{iii}] emission lines — ordered by increasing ionization potential from top to bottom — are shown for the five subsamples, displayed in separate columns. The $W_{\rm 80}$ values are estimated as the flux weighted mean $W_{\rm 80}$ values of spaxels within a radius of 0.5 arcsec centered at the peak or the continuum emission. The mean values (indicated by vertical dotted lines),  standard error, and the number of galaxies are displayed in each panel.    }        
         \label{fig:HistW80}
   \end{figure*}


We use the $W_{\rm 80}$ parameter to trace the velocity dispersion of the gas, which can be associated with turbulence and kinematic disturbances in the nuclear regions.  In Fig.~\ref{fig:HistW80}, we present the nuclear $W_{\rm 80}$ distributions for the H$_2$ S(5)$\lambda6.9091\mu$m, [Fe\:{\sc ii}]$\lambda5.3403\mu$m, [Ar\:{\sc ii}]$\lambda6.9853\mu$m, and [Ar\:{\sc iii}]$\lambda8.9914\mu$m emission lines (from top to bottom) for the  BAT AGN, RS AGN, [Mg\:{\sc v}] AGN, IR AGN and SF subsamples (from left to right, respectively). These values are computed as the flux-weighted mean $W_{\rm 80}$ values measured for spaxels within a 0.5 arcsec radius aperture. 
We observe that, for all AGN samples, the  H$_2$ emission lines consistently present the lowest mean values of $W_{\rm 80}$. Furthermore, the $W_{\rm 80}$ values of the ionized gas emission lines increase from [Fe\:{\sc ii}] through [Ar\:{\sc ii}], reaching the highest values for [Ar\:{\sc iii}], following the order of increasing ionization potential. These higher $W_{\rm 80}$ values indicate enhanced gas turbulence in the more highly ionized regions. For example, the mean difference of $W_{\rm 80}$ between [Fe\:{\sc ii}] and H$_2$, considering all AGN subsamples, is $56 \pm 28$ km\,s$^{-1}$, 
while the mean difference between [Ar\:{\sc iii}] and [Fe\:{\sc ii}] is $115 \pm 35$ km\,s$^{-1}$. On the other hand, the SF subsample shows similar mean $W_{\rm 80}$ values for H$_2$, [Fe\:{\sc ii}], and [Ar\:{\sc ii}], while [Ar\:{\sc iii}] exhibits higher values.   Finally, it is noted that the AGN samples include objects with $W_{\rm 80}$ > 500 km\:s$^{-1}$, which exceed the values expected for motions solely governed by the gravitational potential of galaxies and are commonly used as a criterion to identify outflows in AGN hosts in the local universe \citep[e.g.][]{wylezalek20}. Even smaller values, around $\gtrsim300$ km\:s$^{-1}$, have also been linked to ionized gas outflows in low-luminosity AGN \citep{rogemar23_extended_kin,Gatto24}.  Such high $W_{\rm 80}$ also reflects strong turbulence associated with AGN-driven outflows. In addition, three galaxies in the SF sample--M\:104, IRAS\:19542+1110, and IRAS\:13120$-$5453--exhibit nuclear $W_{\rm 80}$ values exceeding 500 km\,s$^{-1}$. In M\:104, this is observed across all ionized-gas emission lines, while in the other two it is present only in [Ar\,{\sc iii}].

   \begin{figure*}[ht]
   \centering
   \includegraphics[width=0.49\textwidth]{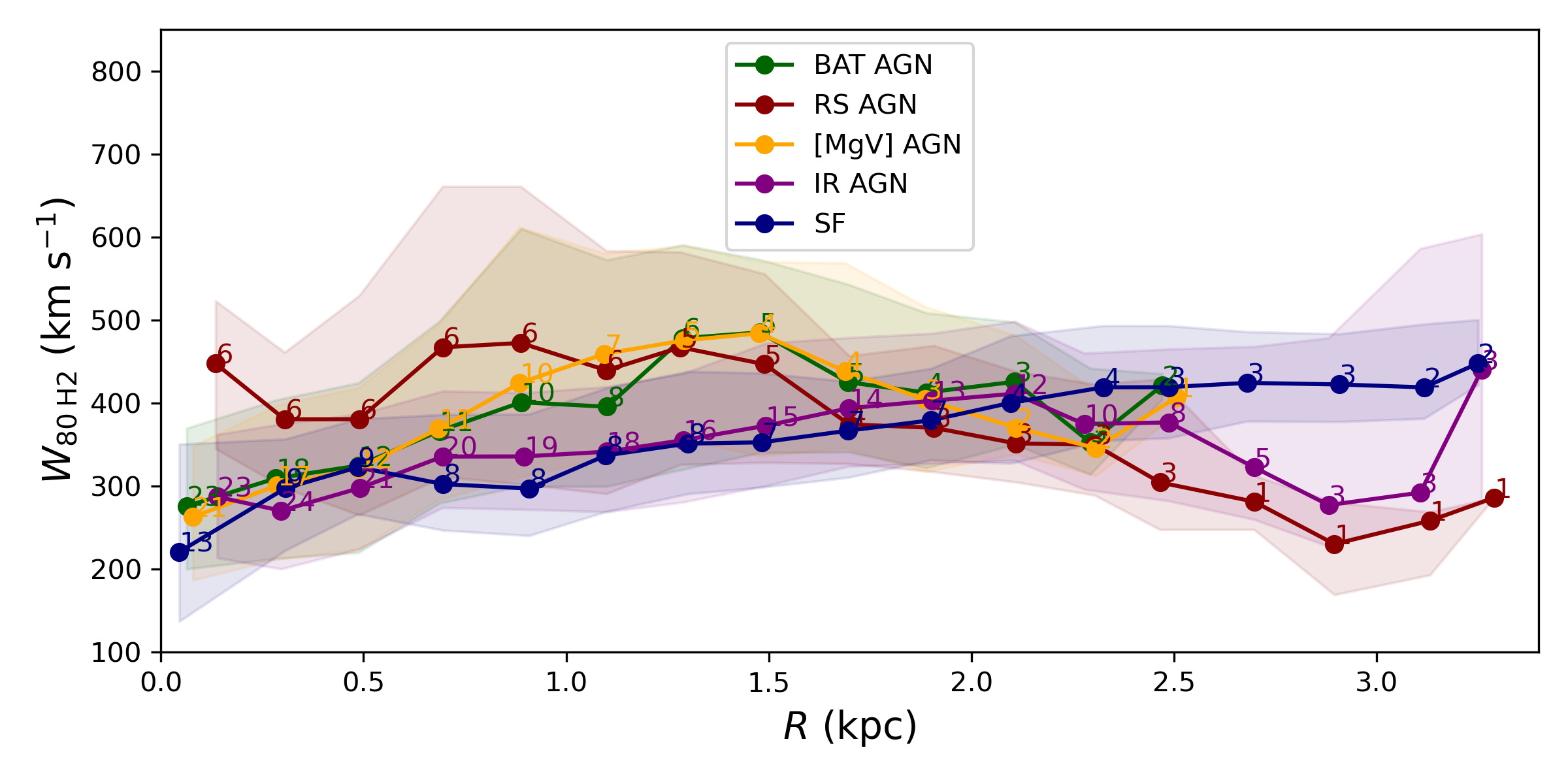}
    \includegraphics[width=0.49\textwidth]{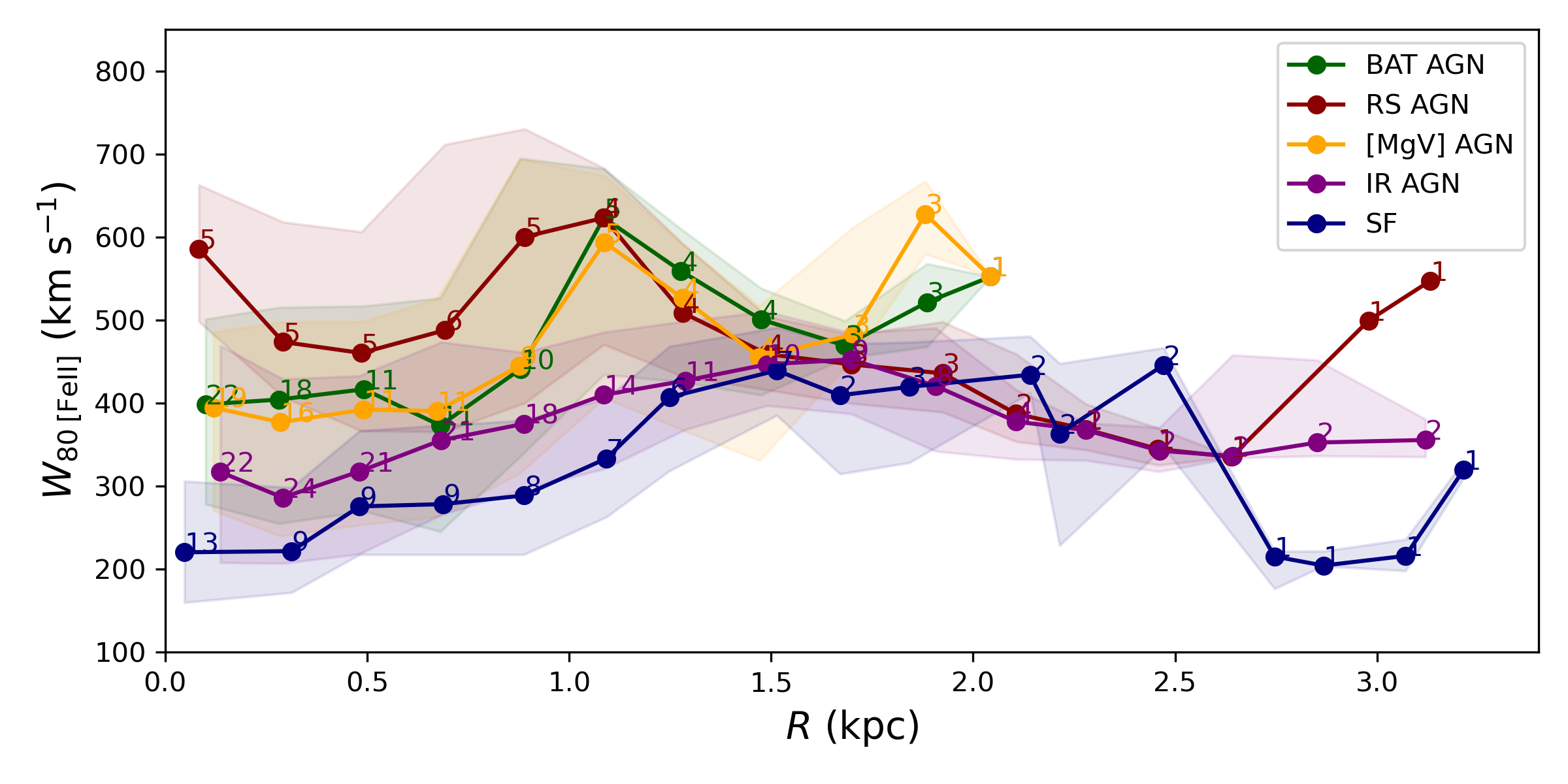}
    \includegraphics[width=0.49\textwidth]{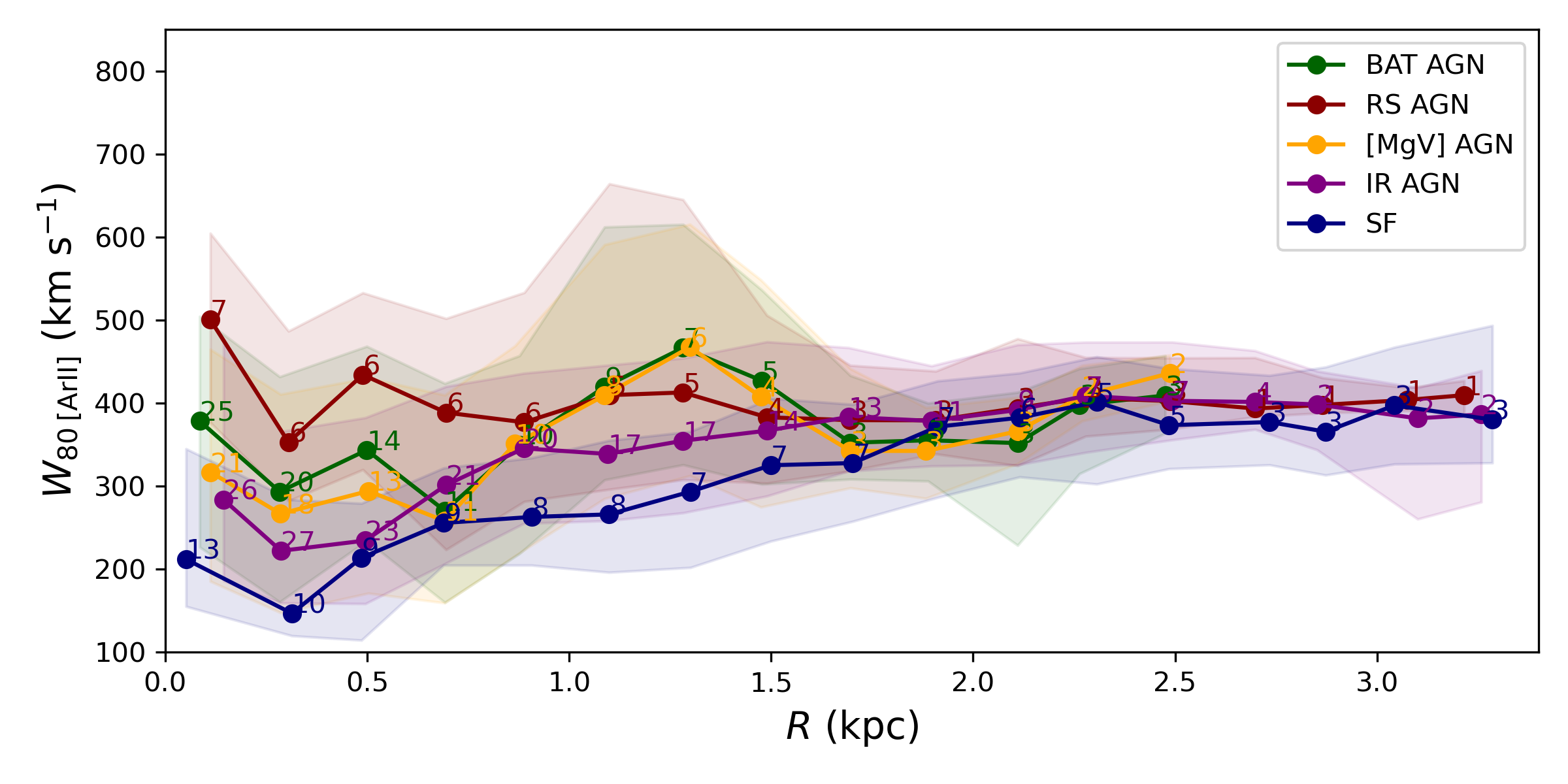}
    \includegraphics[width=0.49\textwidth]{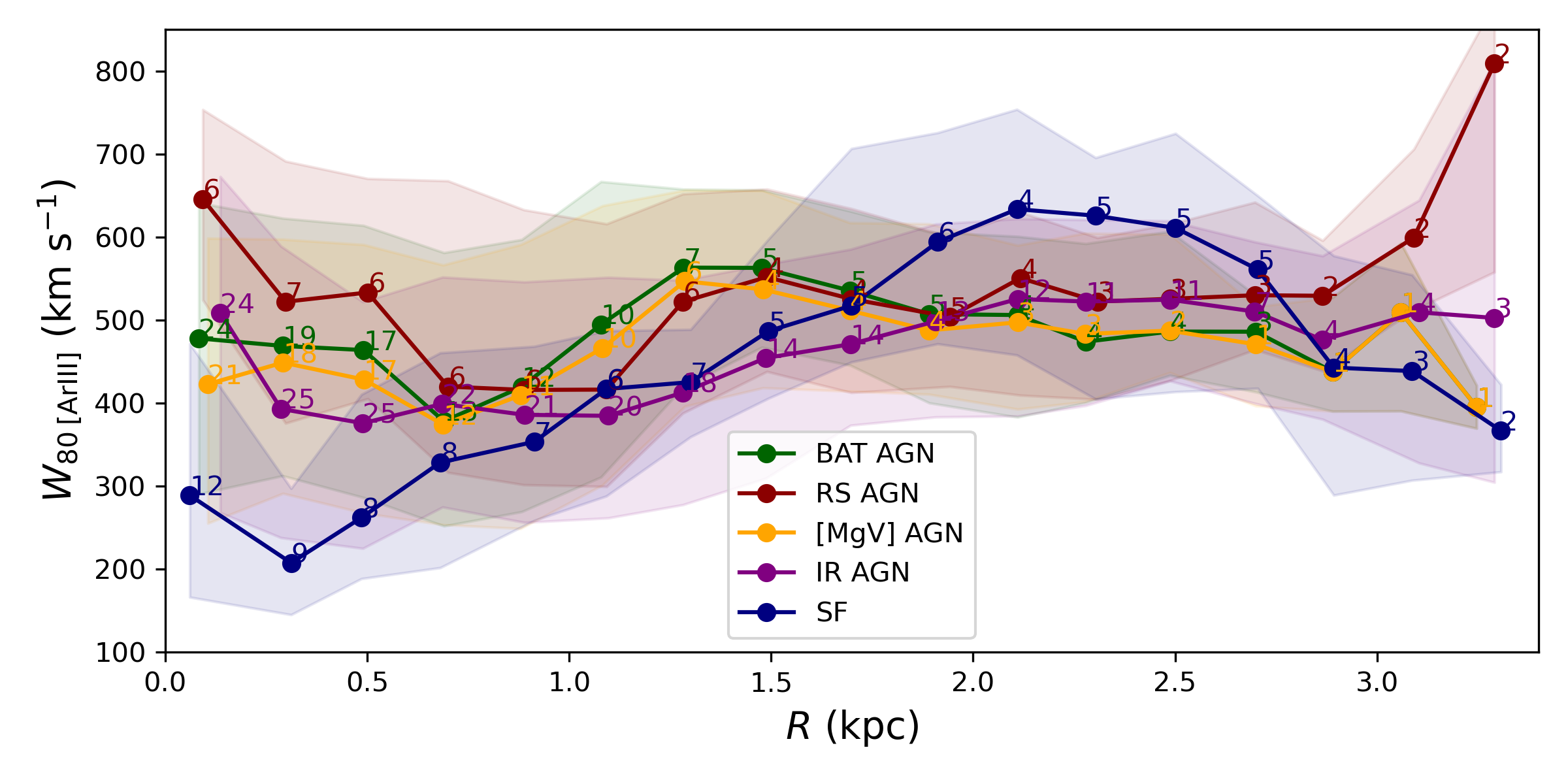}
     \caption{Radial $W_{\rm 80}$ profiles for H$_2$ (top left), [Fe\:{\sc ii}] (top right),  [Ar\:{\sc ii}] (bottom left) and [Ar\:{\sc iii}] (bottom right) for the five subsamples, as indicated by the different colors. These profiles are computed as median values of $W_{\rm 80}$ and distance of the spaxel from the position of the continuum peak, within circular rings of 250 pc width. The numbers next to each point indicate the number of galaxies used to compute it, and the shaded regions represent the range between the 25th and 75th percentiles of the $W_{\rm 80}$ values within each radial bin, illustrating the spread of values observed in each emission line}.
         \label{fig:W80_radial_lines}
   \end{figure*}

In Fig.~\ref{fig:W80_radial_lines} we show $W_{\rm 80}$  radial profiles for the H$_2$ (top left), [Fe\:{\sc ii}] (top right),  [Ar\:{\sc ii}] (bottom left) and [Ar\:{\sc iii}] (bottom right) for the five subsamples. These radial profiles were constructed by computing the median $W_{\rm 80}$ and median radii values within 250 pc radial bins, considering all spaxels in all galaxies in each sample. The radial distances are calculated relative to the position of the peak of the continuum emission at 6.75\:$\mu$m.  The bin size was chosen as a compromise between the angular resolution for more distant objects and the field-of-view size for nearer ones. The bin width is smaller than the PSF radius only for the four most distant objects ($D \gtrsim 300$ Mpc). 
 The number of galaxies used in each bin is indicated next to each data point.   Fig.~\ref{fig:W80_radial_types}  presents the same $W_{\rm 80}$ radial profiles, but grouping the four emission lines for each subsample.

The lowest values of  $W_{\rm 80}$ at most radii are found for the SF sample across all emission lines within distances up to 1.5–2 kpc, beyond which the number of objects with spatial coverage for larger distances becomes small. An exception is seen in [Ar\:{\sc iii}], where the $W_{\rm 80}$ values for SFGs at distances greater than 1 kpc exceed those in AGN; however, only two SFGs show emission in this line beyond 1 kpc.   For the ionized gas lines, there is a trend of decreasing $W_{\rm 80}$ values from the nucleus up to $\sim0.5-1$ kpc, followed by an increase with radius in more distant regions. These variations likely reflect changes in the turbulent motions of the gas across different spatial scales.  This behaviour is less evident in [Fe\:{\sc ii}], where the radial profiles show more monotonically increasing values with the distance from the nucleus, except for the RS AGN. At distances larger than $\sim$2~kpc, the $W_{\rm 80}$ values start to decrease, but the number of objects with measurements at these scales is very reduced. 
For H$_2$, the radial profiles show values increasing with distance from the nucleus for all subsamples, except in the RS AGN, where a similar behavior to that described above for the ionized gas is observed.  Furthermore, the gas velocity dispersion values in our SF sample are significantly higher than those observed in low-$z$ SFGs, which typically exhibit $\sigma \approx 10$--$40$\,km\,s$^{-1}$ or $W_{\rm 80} \approx 25$--$100$\,km\,s$^{-1}$ \citep[e.g.,][]{Epinat2010}. Instead, our measurements are comparable to those observed in local U/LIRGs, which are often attributed to gas outflows \citep[e.g.,][]{Piqueras-Lopez12,Bellocchi13,Arribas14}.

   \begin{figure*}[!ht]
   \centering
   \includegraphics[width=0.99\textwidth]{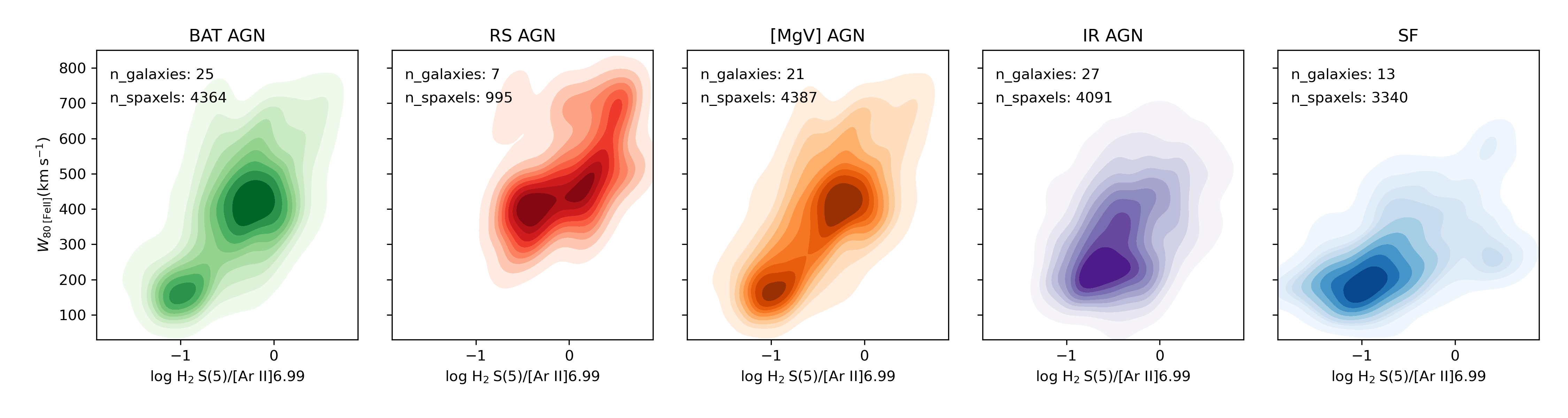}
       \includegraphics[width=0.99\textwidth]{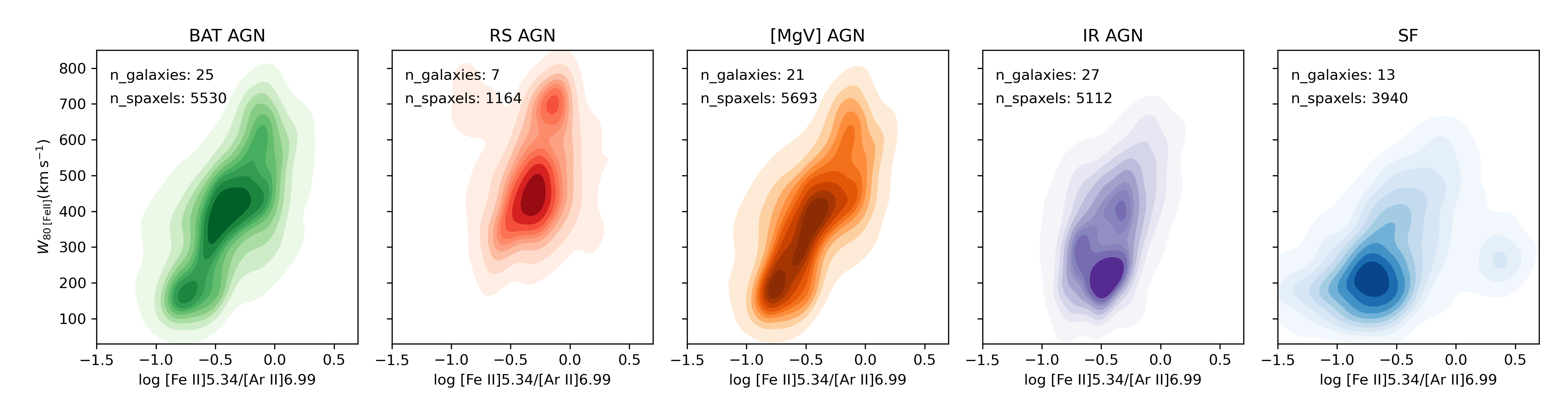}
       \includegraphics[width=0.99\textwidth]{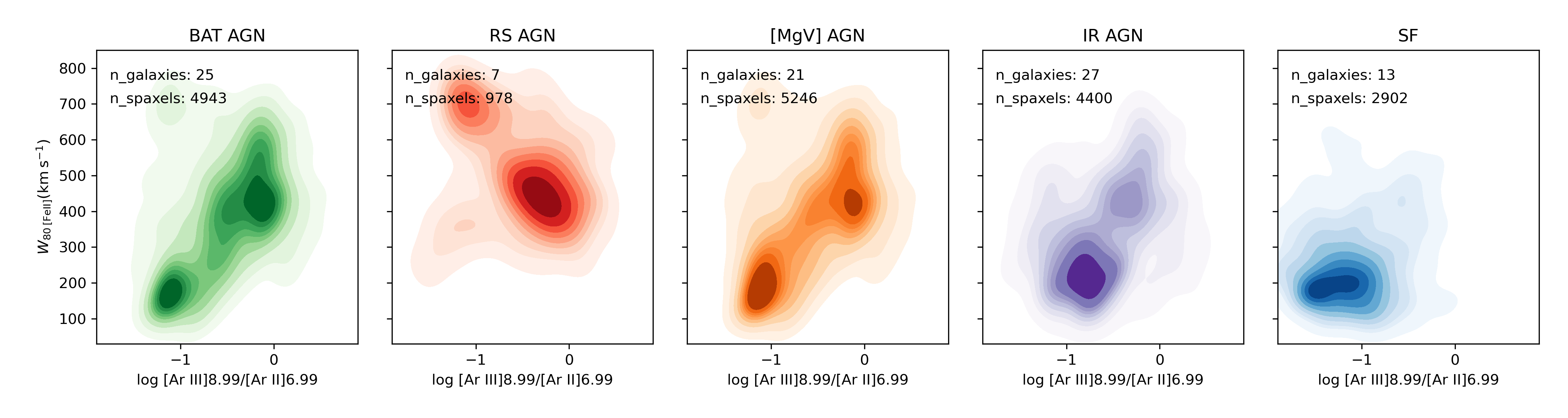}
         \caption{Density maps of $W_{\rm 80}$ for $[\mathrm{Fe\,\textsc{ii}}] \, \lambda 5.3403 \, \mu\mathrm{m}$ emission line as a function of the flux ratios: $ \mathrm{H_2 \, S(5)} \, \lambda 6.9091 \, \mu\mathrm{m}/[\mathrm{Ar\,\textsc{ii}}] \, \lambda 6.9853 \, \mu\mathrm{m} $ (top panels), $ [\mathrm{Fe\,\textsc{ii}}] \, \lambda 5.3403 \, \mu\mathrm{m}/[\mathrm{Ar\,\textsc{ii}}] \, \lambda 6.9853 \, \mu\mathrm{m} $ (middle panels), and $ [\mathrm{Ar\,\textsc{iii}}] \, \lambda 8.9914 \, \mu\mathrm{m}/[\mathrm{Ar\,\textsc{ii}}] \, \lambda 6.9853 \, \mu\mathrm{m} $ (bottom panels).  The contours are spaced at intervals of 10\% in data density.  Results are shown for the different subsamples, as identified in the title of each panel.}              
         \label{fig:W80rat}
   \end{figure*}

Emission-line flux ratios can be used to investigate the origin of gas emission, while the line widths provide insights into the origin of gas turbulence and the behaviors observed in the radial profiles of $W_{\rm 80}$. Fig.~\ref{fig:W80rat} presents $W_{\rm 80}$ density maps for [Fe\:{\sc ii}]$\lambda5.3403\mu$m versus the flux ratios H$_2$ S(5)$\lambda6.9091\mu$m/[Ar\:{\sc ii}]$\lambda6.9853\mu$m (top panels), [Fe\:{\sc ii}]$\lambda5.3403\mu$m/[Ar\:{\sc ii}]$\lambda6.9853\mu$m (central panels) and [Ar\:{\sc iii}]$\lambda8.9914\mu$m/[Ar\:{\sc ii}]$\lambda6.9853\mu$m (bottom panels) for the different subsamples. These maps were constructed using spaxel-based measurements. 

As observed in Fig.~\ref{fig:W80rat}, the SF sample exhibits the highest densities at $W_{\rm 80} \lesssim 300 \, \mathrm{km \, s^{-1}}$, while the AGN samples display higher values, typically ranging between $300$ and $ 600 \, \mathrm{km \, s^{-1}}$, with a secondary clustering in the distribution of points at values similar to those of SFGs, except for the RS AGN. 
The clustering of points with lower values of $W_{\rm 80}$ and line ratio (the secondary clustering) is associated with spaxels located closer to the nucleus, compared to the clustering of points with higher values of these parameters. For regions with $W_{\rm 80} < 300\, \mathrm{km\,s^{-1}}$ (and low line ratios, as seen in Fig.~\ref{fig:W80rat}), the median distance of spaxels from the nucleus is 234, 282, and 340 pc for the BAT, [Mg\,{\sc v}], and IR AGN samples, respectively. For regions with $W_{\rm 80} > 300\, \mathrm{km\,s^{-1}}$, these distances increase to 267, 321, and 534 pc for the same samples. The secondary clustering detected in the AGN samples is probably linked to SF within the MIRI FoV, as observed in several of these galaxies \citep[e.g.][]{Ricci18,Zanchettin24,Cassanta25}. The highest $W_{\rm 80}$  values for the AGN samples reach $800 \, \mathrm{km \, s^{-1}}$, whereas for the SF sample, the maximum values reach up to $600 \, \mathrm{km \, s^{-1}}$.  The BAT and [Mg\,{\sc v}] AGN subsamples show remarkably similar distributions in these plots. This is not surprising, since both are direct tracers of AGN emission, either through the X-rays produced by the central engine or through the high-energy photons required to ionize the coronal gas. For all subsamples, a trend of increasing $ \mathrm{H_2 \, S(5)} \, \lambda 6.9091 \, \mu\mathrm{m}/[\mathrm{Ar\,\textsc{ii}}] \, \lambda 6.9853 \, \mu\mathrm{m} $ and $ [\mathrm{Fe\,\textsc{ii}}] \, \lambda 5.3403 \, \mu\mathrm{m}/[\mathrm{Ar\,\textsc{ii}}] \, \lambda 6.9853 \, \mu\mathrm{m} $ with the line width of $ [\mathrm{Fe\,\textsc{ii}}] $ is observed. However, the SF sample exhibit lower values of flux ratios compared to the AGN samples. A similar trend is observed with the $ [\mathrm{Ar\,\textsc{iii}}]/[\mathrm{Ar\,\textsc{ii}}]$ flux ratio (bottom panels), a diagnostic of the ionization degree of the gas since it involves lines of the same element with different ionization states.  However, this trend is only evident for the BAT,   [Mg\,{\sc v}], and IR AGN samples. For the SF sample, most points are concentrated at low flux ratios and $W_{\rm 80}$ values, with some contours extending to higher values. In contrast, the RS AGN sample exhibits two structures: one with $ W_{\rm 80} < 600 \, \mathrm{km \, s^{-1}} $ spanning a wide range of $ [\mathrm{Ar\,\textsc{iii}}]/[\mathrm{Ar\,\textsc{ii}}] $ ratios, and another with $ W_{\rm 80} > 600 \, \mathrm{km \, s^{-1}} $ and lower flux ratios.

\section{Discussion}

As presented in the previous section, we found that: (i) AGN exhibit higher gas turbulence, as indicated by the emission line widths, compared to SFGs. (ii) There is a trend of increasing gas turbulence with distance from the nucleus, more evident for $ \mathrm{H_2} $ and $ [\mathrm{Fe\,\textsc{ii}}]$ (Fig.~\ref{fig:W80_radial_lines}). RS AGN present more turbulent gas compared to other samples. (iii) There is a correlation between the width of the $ [\mathrm{Fe\,\textsc{ii}}] $ emission line and the line ratios $ \mathrm{H_2 \, S(5)} \, \lambda 6.9091 \, \mu\mathrm{m}/[\mathrm{Ar\,\textsc{ii}}] \, \lambda 6.9853 \, \mu\mathrm{m} $ and $ [\mathrm{Fe\,\textsc{ii}}] \, \lambda 5.3403 \, \mu\mathrm{m}/[\mathrm{Ar\,\textsc{ii}}] \, \lambda 6.9853 \, \mu\mathrm{m} $ for all subsamples  (Fig.~\ref{fig:W80rat}). However, only AGN samples show a trend of increasing $ [\mathrm{Ar\,\textsc{iii}}] \, \lambda 8.9914 \, \mu\mathrm{m}/[\mathrm{Ar\,\textsc{ii}}] \, \lambda 6.9853 \, \mu\mathrm{m} $ with gas turbulence. In this section, we investigate the origin of the observed behavior in the $W_{\rm 80}$ radial profiles and the physical mechanisms responsible for the line emission in our sample.

\subsection{$W_{\rm 80}$ radial profiles}

When gas motions are governed by the gravitational potential, a decrease in velocity dispersion with increasing distance from the nucleus is expected; therefore, the observed $W_{\rm 80}$ radial profiles in our sample cannot be solely explained by virial motions.  Such profiles could be interpreted as being generated by outflows or associated with shocks that enhance the turbulence of the gas \citep[e.g.][]{wylezalek20,ruschel-dutra21,Audibert23,Bessiere24}. Similar increases in velocity dispersion with distance from the nucleus are observed in the inner few hundred parsecs of nearby AGN hosts, in both hot molecular gas (traced by the H$_2$~2.1218~$\mu$m line) and ionized gas (traced by the Br$\gamma$ line), based on near-IR integral field spectroscopy observations of a sample of 31 AGN, in which KDRs are detected in 94\% of the cases for the ionized gas and in 76\% for the hot molecular gas \citep{rogemar23_extended_kin}. A similar result is found using a larger sample of 88 AGN, in which flat or increasing velocity dispersion profiles are observed within the inner 200~pc for obscured, unobscured, and LINER sources, including in the coronal gas traced by the [Si\,\textsc{vi}]~1.9641~$\mu$m emission line \citep{Delaney25}. The origin of the enhanced line widths in these sources is associated with outflows and the shocks they produce \citep[e.g.][]{rogemar21_survey}.  On galaxy-wide scales, AGN hosts exhibit enhanced velocity dispersion relative to normal galaxies, extending out to distances of up to 2 effective radii, as revealed by observations from the Mapping Nearby Galaxies at Apache Point Observatory (MaNGA) survey using the [O\,\textsc{iii}]~$\lambda5007$ emission line \citep{wylezalek20,Alban24}.  Radio‐selected AGN exhibit the largest differences at all radii, indicating that AGN‐driven kinematic perturbations in this population have been active for longer durations than in purely photoionized AGN, consistent with radio emission originating from shocks associated with outflows \citep{Alban24}.  On the other hand, \citet{Kukreti25}, using MaNGA data, find that radio AGN exhibit higher  [O\,\textsc{iii}]~$\lambda5007$ velocity dispersions than optical AGN within the inner 0.5 effective radii, but show lower values at larger radii. These authors suggest that the discrepancy with the results of \citet{Alban24} could be explained by the presence of low-luminosity AGN in their optical AGN sample, which are not included in the \citet{Alban24} sample due to different selection methods. Despite the discrepancies at large distances from the nucleus, the results for the central regions of the galaxies, which are the focus of the present work, indicate that radio AGN exhibit higher gas turbulence.

Recently, \citet{Marconcini25_nat} modeled the ionized gas kinematics of a sample of 10 galaxies observed with the Multi Unit Spectroscopic Explorer (MUSE), using the MOKA$^{\rm 3D}$ tool \citep{Marconcini23}. They found that ionized outflows exhibit constant or slightly decreasing velocities within the inner 1~kpc from the nucleus, followed by a rapid increase, very similar to the behavior observed in our sample. \citet{Marconcini25_nat} argue that the increase in outflow velocity, relative to the galaxy escape velocity, is observed in the transition region, where a momentum-driven phase shifts to an energy-conserving phase beyond 1~kpc, with the outflow expanding as an isothermal, consistent with theoretical predictions of AGN outflows \citep[e.g.][]{King15}. Thus, the observed behavior of the $W_{\rm 80}$ radial profiles in our sample, when compared with previous results, suggests that they are associated with AGN- or star-formation-driven energy-conserving winds capable of escaping the bulge of the galaxy, in both low-ionization and warm molecular gas phases.  These findings indicate that the enhanced gas turbulence observed in our sample may play a significant role in the evolution of the host galaxies. However, a comprehensive analysis of the gas kinematics, including detailed modeling of individual sources, is required to fully characterize the outflow properties and assess their potential impact on SF.

\subsection{Origin of the emission}

The emission of [Fe\:{\sc ii}] is highly sensitive to shocks, increasing significantly when shocks release iron from dust grains \citep{Oliva01,Hashimoto11,Aliste25}. The emission of H$_2$ lines can also be enhanced in regions of shocked gas \citep{Hollenbach89,Guillard09,Kristensen23,Appleton23,Godard24,Zanchettin25}. The $ [\mathrm{Ar\,\textsc{iii}}] \, \lambda 8.9914 \, \mu\mathrm{m}/[\mathrm{Ar\,\textsc{ii}}] \, \lambda 6.9853 \, \mu\mathrm{m} $ flux ratio can be used to map the ionization degree and the intensity of the incident radiation field in photoionized gas regions, as their parent ions have significantly different ionization potentials -- 15.8 eV and 27.6 eV for Ar\:{\sc ii}  and   for Ar\:{\sc iii}, respectively. 
 The observed trend of increasing $W_{\rm 80}$ with increasing values of the ${\rm H_2 \, S(5)}$/[Ar\,{\sc ii}]$\lambda6.9853\:\mu$m and [Fe\,{\sc ii}]$\lambda5.3403\:\mu$m/[Ar\,{\sc ii}]$\lambda6.9853\:\mu$m (first two rows of Fig.~\ref{fig:W80rat}) could be interpreted as an additional contribution from shocks to the excitation of H$_2$ and [Fe\:{\sc ii}] emission lines. On the other hand, an increase in the [Ar\,{\sc iii}]$\lambda8.9914\:\mu$m/[Ar\,{\sc ii}]$\lambda6.9853\:\mu$m ratio for the larger $W_{\rm 80}$, seen for the AGN samples, might suggest that although part of the gas is in the outflow, it remains predominantly photoionized by the central AGN.

   \begin{figure*}
   \centering
   \includegraphics[width=0.99\textwidth]{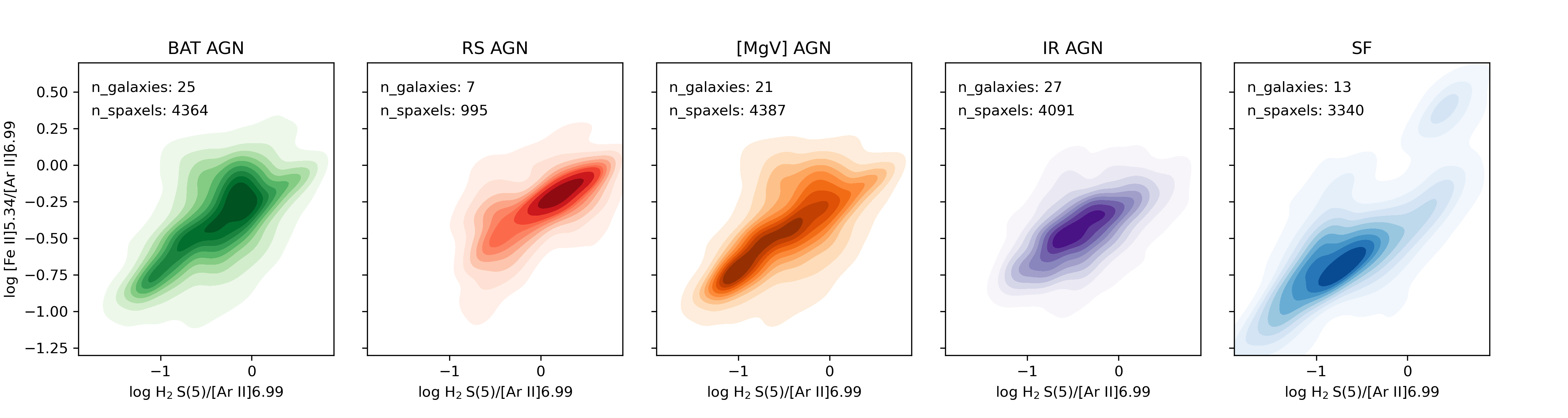}
      \includegraphics[width=0.99\textwidth]{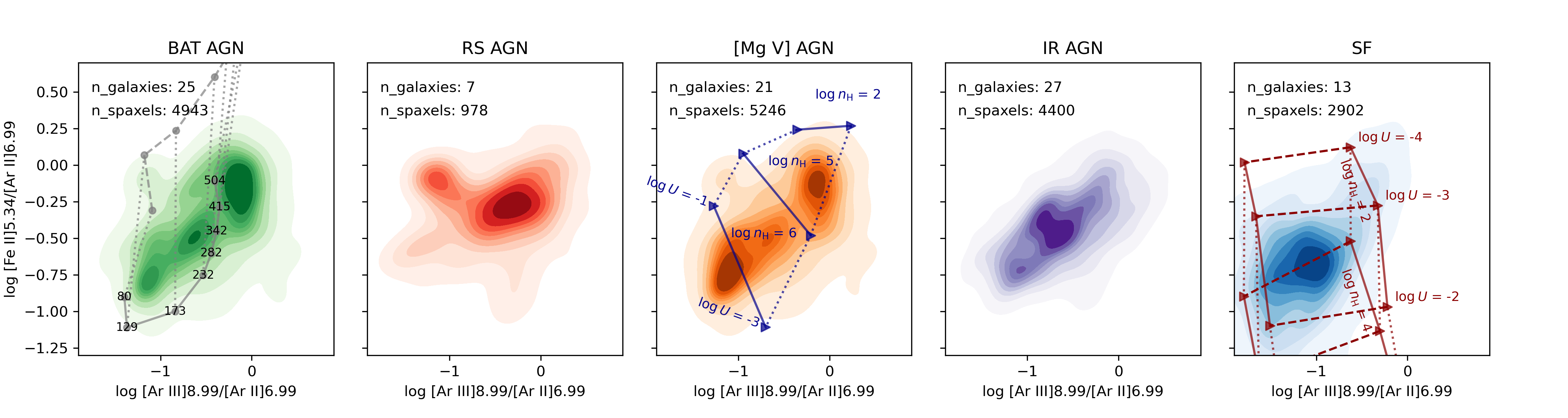}
         \caption{Density plots for the [Fe\,{\sc ii}]$\lambda5.3403\:\mu$m/[Ar\,{\sc ii}]$\lambda6.9853\:\mu$m vs. ${\rm H_2 \, S(5)}$/[Ar\,{\sc ii}]$\lambda6.9853\:\mu$m (top panels) and [Fe\,{\sc ii}]$\lambda5.3403\:\mu$m/[Ar\,{\sc ii}]$\lambda6.9853\:\mu$m vs. [Ar\,{\sc iii}]$\lambda8.9914\:\mu$m/[Ar\,{\sc ii}]$\lambda6.9853\:\mu$m (bottom panels). The dark red lines (bottom-right panel) show predictions from photoionization models of star-forming regions, assuming stellar clusters aged 6 and 7 Myr. Solid lines indicate different gas densities, dotted lines correspond to varying ionization parameters (as labeled in the figure), and the predictions for both cluster ages are connected by dashed lines. Similarly,  results from AGN photoionization models are shown as dark blue lines (bottom-central panel). The gray lines and filled circles (bottom-left panel) are predictions of fast shock models from \citet{Pereira-Santaella24b} using the \textsc{mappings v} code \citep{Sutherland17} for $\log R_{\rm P}=6$ (dashed) and $\log R_{\rm P}=8$ (continuous). The numbers indicate the shock velocities in km\:s$^{-1}$.
 }             
         \label{fig:dens_ratios}
   \end{figure*}

   \begin{figure*}
   \centering
           \includegraphics[width=0.99\textwidth]{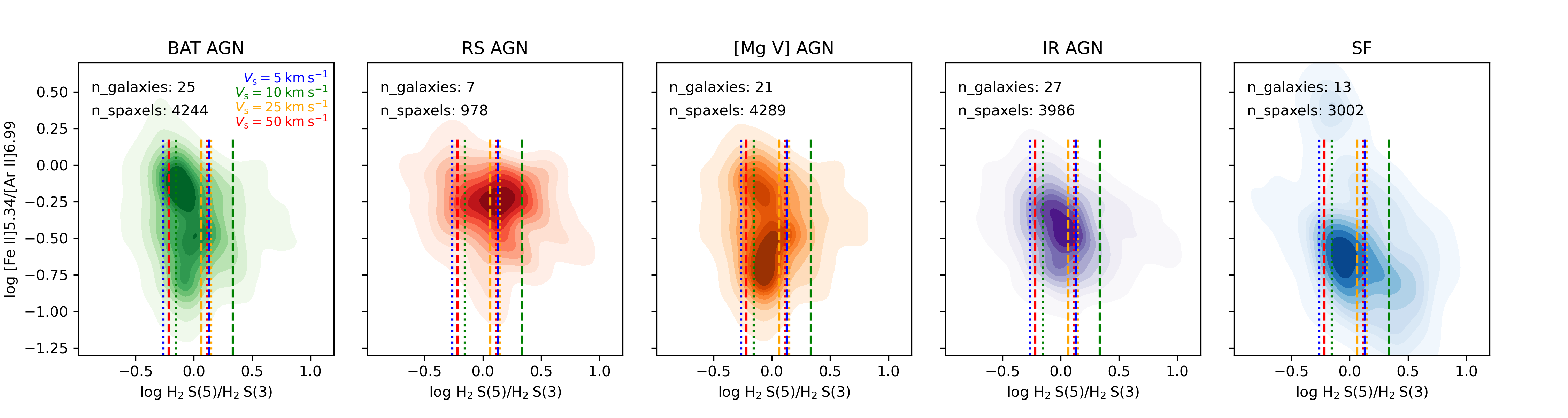}
        \includegraphics[width=0.99\textwidth]{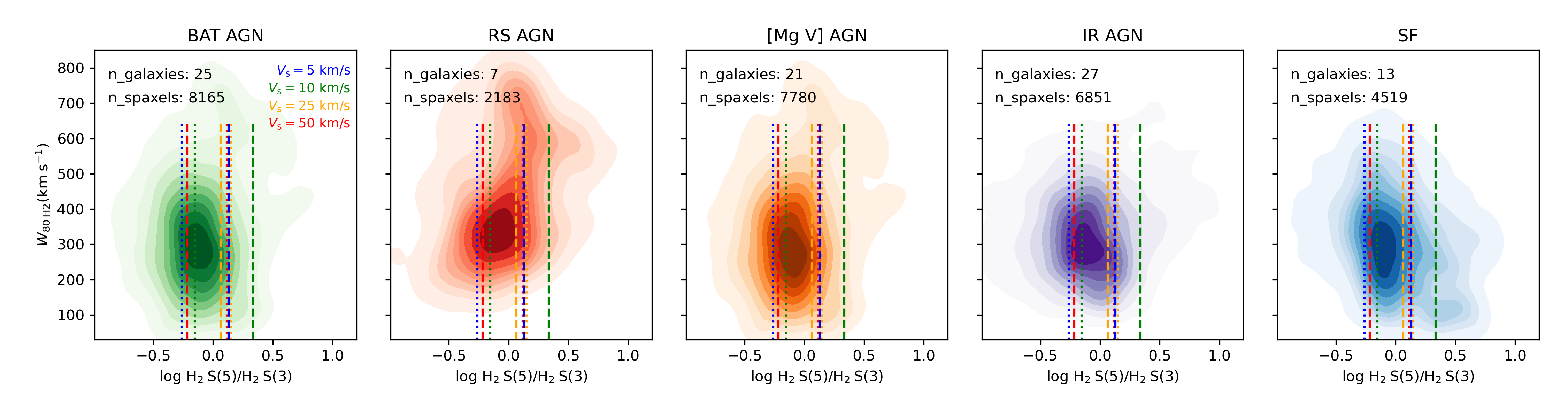}
         \caption{Density plots for [Fe\,{\sc ii}]$\lambda5.3403\:\mu$m/[Ar\,{\sc ii}]$\lambda6.9853\:\mu$m vs. H$_2$\:S(5)/H$_2$\:S(3) (top) and $W_{\rm 80}$ for the H$_2$ S(5) line vs.  H$_2$\:S(5)/H$_2$\:S(3) (bottom)  for all subsamples.  The vertical lines represent mean H$_2$\:S(5)/H$_2$\:S(3) predictions of the shock models from \citet{Kristensen23} for velocities of 5, 10, 25 and 50 km\:s$^{-1}$ (different colors) and densities of $n_H=10^3$ (dotted lines) and   $n_H=10^6$ cm$^{-1}$ (dashed lines). 
 }             
         \label{fig:dens_ratios_h2}
   \end{figure*}

To further investigate the origin of gas turbulence and the trends observed in emission-line flux ratios, we present in Fig.~\ref{fig:dens_ratios} density plots for the [Fe\,{\sc ii}]$\lambda5.3403\:\mu$m/[Ar\,{\sc ii}]$\lambda6.9853\:\mu$m vs. ${\rm H_2 \, S(5)}$/[Ar\,{\sc ii}]$\lambda6.9853\:\mu$m (top panels) and [Fe\,{\sc ii}]$\lambda5.3403\:\mu$m/[Ar\,{\sc ii}]$\lambda6.9853\:\mu$m vs. [Ar\,{\sc iii}]$\lambda8.9914\:\mu$m/[Ar\,{\sc ii}]$\lambda6.9853\:\mu$m  (bottom panels), for all subsamples. A strong correlation is observed for the first pair of intensity ratios in all samples,  with Pearson correlation coefficients ranging from 0.53 to 0.76.   As [Ar\,\textsc{ii}] is predominantly produced by photoionization, while [Fe\,\textsc{ii}] and H$_2$ are more sensitive to shocks, the observed correlation suggests that these latter lines trace shock-excited gas, particularly at larger distances from the nucleus. Moreover, the increase in line ratios with line widths (Fig.~\ref{fig:W80rat}) provides strong evidence that the enhanced gas turbulence in these regions is driven by shocks. 
 A much weaker trend is observed between [Fe\:{\sc ii}]$\lambda5.3403\:\mu$m/[Ar\:{\sc ii}]$\lambda6.9853\:\mu$m and [Ar\:{\sc iii}]$\lambda8.9914\:\mu$m/[Ar\:{\sc ii}]$\lambda6.9853\:\mu$m, and only for the AGN samples (excluding the RS AGN). No clear relationship is observed between these line ratios for the SF subsample, where argon emission is attributed to SF. Since the [Fe\,{\sc ii}] and H$_2$ fluxes increase relative to [Ar\,{\sc ii}] emission and also increase with line width, this suggests that both [Fe\,{\sc ii}] and H$_2$ exhibit enhanced emission due to shocks, which could be associated with stellar winds and mergers. This is consistent with the fact that the SF subsample is predominantly composed of U/LIRGs, which are known for intense SF activity and their presence in merger systems. Similarly, no clear relationship is observed between these line ratios for the RS AGN sample. In this case, the possible origin of the shocked [Fe\,{\sc ii}] and H$_2$ gas may be associated with radio emission, stellar winds, or interactions in the U/LIRGs of this subsample.

We can compare the observed emission-line ratios with predictions from shock and photoionization models.  The results from fast shock models by \citet{Pereira-Santaella24b}, using the \textsc{mappings v} code \citep{Dopita96,Sutherland17},  are shown overlaid as gray lines in the bottom-left panel of Fig.~\ref{fig:dens_ratios}.  A short description of these models are presented in Appendix~\ref{PhotModels}. Although these authors produced an extensive grid of models (see their work for more details), we only show a subset in the figure to avoid overcrowding. The displayed models assume solar metallicity, shock velocities ($v_{\rm s}$) ranging from 80 to 500~km\,s$^{-1}$, and ram pressure parameters defined as $R_{\rm P} = \frac{n_{\rm H}}{\text{cm}^{-3}} \times \left(\frac{v_s}{\text{km/s}}\right)^2$, with values of $10^6$ and $10^8$, where $n_{\rm H}$ is the gas volume density. We also constructed extensive photoionization model grids for AGN and SF using the \textsc{cloudy} code \citep{Ferland17}, as detailed in Appendix~\ref{PhotModels}.  The SF models are shown as dark red lines in the bottom-right panel of Fig.~\ref{fig:dens_ratios}. We present predictions for gas densities of $\log(n_{\rm H}/{\rm cm^{-3}}) = 2$ and $4$, and ionization parameters of $\log U = -4$, $-3$, and $-2$, based on Spectral Energy Distributions (SEDs) generated with the \textsc{Starburst99} code \citep{Leitherer99} for stellar clusters aged 6 Myr (yielding higher [Ar\,{\sc iii}]/[Ar\,{\sc ii}] ratios) and 7 Myr (lower ratios). These models are able to reproduce the lower line ratios, supporting a interpretation that they originate from gas photoionized by young stellar populations.  The predictions from AGN photoionization models are shown as blue lines in the bottom-central panel of Fig.~\ref{fig:dens_ratios}, for densities of $\log(n_{\rm H}/{\rm cm^{-3}}) = 2$, $5$, and $6$, and ionization parameters of $\log U = -1$ and $-3$. These models are able to reproduce the highest line ratios, including the extreme [Ar\,{\sc iii}]/[Ar\,{\sc ii}] values observed, which are not well matched by the shock models.

As shown, the observed [Fe\,{\sc ii}]$\lambda5.3403\:\mu$m/[Ar\,{\sc ii}]$\lambda6.9853\:\mu$m flux ratios align well with the values predicted by shock models. In contrast, the [Ar\,{\sc iii}]~$\lambda8.9914\:\mu$m/[Ar\,{\sc ii}]~$\lambda6.9853\:\mu$m ratios exhibit values higher than those predicted by shock models, but are well reproduced by AGN photoionization models. In addition, the lowest values of both ratios are consistent with predictions from SF photoionization models.

 Fig.~\ref{fig:dens_ratios_h2}  displays density plots of [Fe\,{\sc ii}]$\lambda5.3403\:\mu$m/[Ar\,{\sc ii}]$\lambda6.9853\:\mu$m  and $W_{\rm 80}$ of the H$_2$\:S(5) line against the H$_2$\:S(5)/H$_2$\:S(3) line ratio, which is a tracer of the H$_2$ temperature and is enhanced in shock dominated-regions \citep[e.g.][]{zakamska10,Pereira-Santaella14,togi16,Dan25,rogemar25_jwst}. The observed values for this ratio are consistent with those predicted by low velocity shock models (2--90\:km\:s$^{-1}$), as indicated by the vertical lines representing predictions by the models from \citet{Kristensen23}.  However, no evident correlation is observed. This indicates that if the H$_2$ emission is produced by shocks, these are not the same shocks responsible for the [Fe\:{\sc ii}] emission \citep{Hollenbach89,Mouri00,allen08}. This is expected, as the shock waves produced by fast shocks with velocities of ${\rm 200-300 km\:s^{-1} }$ necessary to  generate ionized gas emission would lead to the dissociation of H$_2$ molecules. On the other hand, the tight correlation observed between the [Fe\:{\sc ii}] and H$_2$ emission (top row of Fig.~\ref{fig:dens_ratios}) indicates that they share the same physical origin. 
 A possible interpretation for this correlation is that the H$_2$ emission originates in the post-shock gas, where the molecules reform after being dissociated by the fast shock \citep{Guillard09,richings18a,richings18b}. Another explanation is the excitation of H$_2$ in the molecular shock precursor, as observed in NGC\:7319. In this case, the ionized gas is significantly more turbulent than the molecular gas, with $\sigma\approx300\:{\rm km\:s^{-1}}$ for the ionized gas and $\sigma\approx150\:{\rm km\:s^{-1}}$ for the H$_2$, in regions co-spatial with a radio hotspot \citep{Pereira-Santaella22}. Although the W$_{\rm 80}$ values for H$_2$ are typically lower than those for the ionized gas (Fig.~\ref{fig:HistW80}), the radial profiles of W$_{\rm 80}$ for H$_2$ and [Fe\:{\sc ii}] show similar trends, with values increasing from the nucleus outward, suggesting that the first interpretation is more likely.

 The analysis of emission line ratios, gas kinematics, and comparisons with shock and photoionization models indicates that shocks driven by outflows and jets are a key factor in producing the [Fe\:{\sc ii}] and H$_2$ emission within the inner few kpc of AGN host galaxies. The [Fe\:{\sc ii}] emission arises from partially ionized zones situated beyond the main hydrogen ionization front in narrow-line region clouds, forming a transition between fully ionized and neutral gas \citep{Forbes93,Simpson96}, while the mid-IR H$_2$ lines trace warm molecular gas at temperatures of a few hundred Kelvin \citep{Pereira-Santaella14,togi16}.  Our results suggest that turbulence produced in the ISM by outflows and/or radio jets may constitute an important mechanism of maintenance-mode AGN feedback, as it prevents the gas from efficiently cooling and forming stars, thus regulating star formation.

\section{Conclusions}

We have used archival JWST MIRI/MRS observations of a sample consisting of 54 galaxies at $z<0.1$ to investigate the origin of the warm molecular and low-ionization gas emission. The sample includes SFGs and AGN hosts, with AGN selected based on their X-ray, radio, and coronal line emissions. We investigate the origin of the emission and turbulence of the gas from flux measurements and velocity dispersion, parameterized by the $W_{\rm 80}$ parameter, for  H$_2$ S(5)$\lambda6.9091\mu$m, [Ar\:{\sc ii}]$\lambda6.9853\mu$m, [Fe\:{\sc ii}]$\lambda5.3403\mu$m, [Ar\:{\sc iii}]$\lambda8.9914\mu$m, and [Mg\:{\sc v}]$\lambda5.6098\mu$m  emission lines. Our main conclusions are the following:

\begin{itemize}
    \item  AGN exhibit broader emission lines than SFGs, with the largest velocity dispersions observed for radio-strong AGN. The H$_2$ gas is less turbulent compared to the ionized gas for all the subsamples studied, followed by [Ar\:{\sc ii}], [Ar\:{\sc iii}], and [Fe\:{\sc ii}]. Among the galaxies with coronal emission, the coronal gas shows higher velocity dispersion values compared to lower-ionization lines, suggesting stratified emission clouds. 
   
    \item The velocity dispersion of molecular and low-ionization gas increases with distance from the nucleus, particularly for the emission lines of H$_2$ and [Fe\:{\sc ii}], which is contrary to what is expected from gravitationally dominated motions. Additionally, the high values of $W_{\rm 80}$ indicate the presence of outflows in AGN hosts, while in SF, the $W_{\rm 80}$ values are lower than those for AGN, yet still require an additional component, such as shock-heated gas emission from stellar winds or galaxy interactions.
  
    \item There is a strong correlation between the $W_{\rm 80}$ parameter and the line ratios H$_2$ S(5)$\lambda6.9091\mu$m/[Ar\:{\sc ii}]$\lambda6.9853\mu$m and [Fe\:{\sc ii}]$\lambda5.3403\mu$m/[Ar\:{\sc ii}]$\lambda6.9853\mu$m, as well as between these two ratios themselves. This indicates that the [Fe\:{\sc ii}] and H$_2$ emissions originate from linked physical processes. Since H$_2$ and [Fe\:{\sc ii}] emissions are enhanced in shocked gas, these correlations provide additional evidence that shocks play an important role in the observed emission in our sample.
    
    \item The lowest $W_{\rm 80}$ values ($<$300 km\,s$^{-1}$) across all samples are associated with SF, presenting line ratios consistent with predictions from SF photoionization models for gas densities in the range $10^{2}$–$10^{4}$ cm$^{-3}$. As gas turbulence increases, the contribution from shocks becomes significant, with [Fe\:{\sc ii}]$\lambda5.3403\mu$m/[Ar\:{\sc ii}]$\lambda6.9853\mu$m intensity ratios consistent with predictions from fast shock models, assuming shock velocities in the range of $\sim100$ to $\sim300$ km\,s$^{-1}$ and gas densities between $10^{3}$ and $10^{4}$ cm$^{-3}$. Some contribution from AGN photoionization may be important, particularly for the highest [Ar\,{\sc iii}]~$\lambda8.9914\:\mu$m/[Ar\,{\sc ii}]~$\lambda6.9853\:\mu$m ratios, which are consistent with AGN photoionization models for gas densities of $10^{2}$ and $10^{5}$ cm$^{-3}$.

\item Although the observed H$_2$\,S(5)/H$_2$\,S(3) flux line ratios are consistent with predictions of slow shock models, there is no trend between this line ratio and the H$_2$ velocity dispersion and [Fe\:{\sc ii}]$\lambda5.3403\mu$m/[Ar\:{\sc ii}]$\lambda6.9853\mu$m. 
 These results, along with the similar trends observed in the gas turbulence for [Fe\:{\sc ii}] and H$_2$, increasing from the nucleus outwards, leads to the interpretation that the H$_2$ emission could be produced by molecules reforming in the post-shock region.   

\item We find similarities in line widths and line ratios between the BAT and [Mg\,{\sc v}] AGN subsamples, consistent with the fact that both are direct tracers of AGN emission, whether through the X-rays generated by the central engine or the high-energy photons necessary to ionize the coronal gas. Similarly, among the AGN subsamples, the IR AGN subsample shows line widths and ratios more closely resembling those of the SFGs, consistent with many IR AGN being U/LIRGs that exhibit intense star formation.

\end{itemize}

Our analysis reveals that high turbulence is a pervasive feature present in all the subsamples studied. This turbulence, driven by AGN outflows/jets and/or stellar winds, plays a crucial role in influencing the kinematics and physical state of the ISM in these galaxies. The widespread presence of such energetic feedback mechanisms underscores their importance in galaxy evolution, affecting gas dynamics and potentially regulating SF across diverse environments.

\begin{acknowledgements}

The authors are grateful to the reviewer for the valuable and very constructive suggestions and comments, which greatly contributed to improving our manuscript.  
The data were obtained from the Mikulski Archive for Space Telescopes (MAST) at the Space Telescope Science Institute (STScI), which is operated by the Association of Universities for Research in Astronomy, Inc., under NASA contract NAS 5-03127 for JWST.  The complete dataset can be accessed at the MAST portal, through the DOI \href{https://doi.org/10.17909/zqhf-wg84}{10.17909/zqhf-wg84}.  
RAR acknowledges the support from Conselho Nacional de Desenvolvimento Cient\'ifico e Tecnol\'ogico (CNPq; Proj. 303450/2022-3, 403398/2023-1, \& 441722/2023-7) and Coordena\c c\~ao de Aperfei\c coamento de Pessoal de N\'ivel Superior (CAPES;  Proj. 88887.894973/2023-00).
CRA and AA acknowledge support from the Agencia Estatal de Investigaci\'on of the Ministerio de Ciencia, Innovaci\'on y Universidades (MCIU/AEI) under the grant ``Tracking active galactic nuclei feedback from parsec to kiloparsec scales'', with reference PID2022$-$141105NB$-$I00 and the European Regional Development Fund (ERDF). 
MPS acknowledges support under grants RYC2021-033094-I, CNS2023-145506, and PID2023-146667NB-I00 funded by MCIN/AEI/10.13039/501100011033 and the European Union NextGenerationEU/PRTR. 
AA acknowledges funding from the European Union grant WIDERA ExGal-Twin, GA 101158446. 
EB acknowledges support from the Spanish grants PID2022-138621NB-I00 and PID2021-123417OB-I00, funded by MCIN/AEI/10.13039/501100011033/FEDER, EU.
FE and SGB acknowledge support from the Spanish grant PID2022-138560NB-I00, funded by MCIN/AEI/10.13039/501100011033/FEDER, EU. 
AJB acknowledges funding from the “FirstGalaxies” Advanced Grant from the European Research Council (ERC) under the European Union’s Horizon 2020 research and innovation program (Grant agreement No. 789056).  
AAH and MVM acknowledge support from grant PID2021-124665NB-I00 funded by the 
Spanish Ministry of Science and Innovation and the State Agency of 
Research MCIN/AEI/10.13039/501100011033 and ERDF A way of making Europe. 
EB acknowledges support from the Spanish grants PID2022-138621NB-I00 and PID2021-123417OB-I00, funded by MCIN/AEI/10.13039/501100011033/FEDER, EU. 
OGM acknowledges financial support from the UNAM PAPIIT project IN109123 and SECIHTI Cienca de Frontera proyect CF-2023-G100.  
EKSH and LZ acknowledge grant support from the Space Telescope Science Institute (ID: JWST-GO-01670). 
This research has made use of the NASA/IPAC Extragalactic Database (NED),
which is operated by the Jet Propulsion Laboratory, California Institute of Technology,
under contract with the National Aeronautics and Space Administration. The AI tool ChatGPT (GPT-4.5) was used to assist in debugging the code used to produce the figures and to refine the wording of some sentences in the manuscript. This research made use of Astropy, a community-developed core Python package for Astronomy \citep{Astropy13,Astropy22}. 

\end{acknowledgements}

%
%


\bibliography{riffel}

 \appendix
 \section{Observational proposals}
 
Our sample consists of 54 galaxies, listed in Table~\ref{tab:sample}, along with details of the observational proposals.  Our sample includes representatives from the The Great Observatories All-Sky LIRG Survey \citep[GOALS; e.g.][]{Armus09,U22,Evans22,Buiten24}, The Galaxy Activity, Torus, and Outflow Survey \citep[GATOS; e.g.][]{Garcia-Burillo21,Alonso-Herrero21,Davies24,Garcia-Bernete24b}, Mid-IR Characterization Of Nearby Iconic galaxy Centers \citep[MICONIC; e.g.][]{Alonso-Herrero24,Alonso-Herrero25,HermosaMunoz25}, Mid-InfraRed Activity of Circumnuclear Line Emission  \citep[MIRACLE; e.g.][]{Marconcini25,Ceci25} projects, among others.

\begin{table*}[ht]
    \caption[]{The sample.  }
    \label{tab:sample}
\centering
\small
\begin{tabular}{|l|c|c|c|c|c|}
\hline
Galaxy & RA  & Dec & Redshift & PID & PI  \\
  &    (HH:MM:SS) &  (DD:MM:SS) &  &  &   \\
\hline
Arp220 & 15:34:57.3 & 23:30:11.4 & 0.0184 & 1267 & Dicken, D \\
Centaurus A & 13:25:27.6 & -43:01:08.8 & 0.0018 & 1269 & Luetzgendorf, N \\
Cygnus A & 19:59:28.4 & 40:44:02.1 & 0.0562 & 4065 & Ogle, P \\
ESO137-G034 & 16:35:14.0 & -58:04:47.9 & 0.0095 & 1670 & Shimizu, T \\
ESO420-G13 & 04:13:49.7 & -32:00:25.2 & 0.0119 & 1875 & Fernandez Ontiveros, J \\
IC5063 & 20:52:02.3 & -57:04:07.6 & 0.0113 & 2004 & Dasyra, K \\
IIZw96 & 20:57:24.4 & 17:07:39.7 & 0.0361 & 1328 & Armus, L \\
IRAS05189-2524 & 05:21:01.4 & -25:21:45.4 & 0.0441 & 3368 & Armus, L \\
IRAS07251-0248 & 07:27:37.6 & -02:54:54.2 & 0.0876 & 3368 & Armus, L \\
IRAS09022-3615 & 09:04:12.7 & -36:27:01.7 & 0.0596 & 3368 & Armus, L \\
IRAS09111-1007 & 09:13:36.5 & -10:19:30.1 & 0.0541 & 3368 & Armus, L \\
IRAS10565+2448 & 10:59:18.1 & 24:32:34.3 & 0.0431 & 3368 & Armus, L \\
IRAS-5453 & 13:15:06.3 & -55:09:22.7 & 0.0308 & 3368 & Armus, L \\
IRAS14348-1447 & 14:37:38.4 & -15:00:21.3 & 0.0823 & 3368 & Armus, L \\
IRAS15250+3608 & 15:26:59.4 & 35:58:37.5 & 0.0552 & 3368 & Armus, L \\
IRAS19297-0406 & 19:32:22.3 & -04:00:01.0 & 0.0857 & 3368 & Armus, L \\
IRAS19542+1110 & 19:56:35.8 & 11:19:04.4 & 0.0650 & 3368 & Armus, L \\
IRAS20551-4250 & 20:58:26.8 & -42:39:00.3 & 0.0430 & 3368 & Armus, L \\
IRAS22491-1808 & 22:51:49.3 & -17:52:24.0 & 0.0777 & 3368 & Armus, L \\
IRAS23128-5919 & 23:15:46.7 & -59:03:11.1 & 0.0446 & 3368 & Armus, L \\
IRASF01364-1042 & 01:38:52.8 & -10:27:11.8 & 0.0482 & 1717 & U, V \\
IRASF08572+3915NW & 09:00:25.4 & 39:03:54.2 & 0.0582 & 3869 & Veilleux, S \\
IRASF14378-3651 & 14:40:59.0 & -37:04:31.9 & 0.0676 & 3869 & Veilleux, S \\
IRASF23365+3604 & 23:39:01.3 & 36:21:08.3 & 0.0645 & 3869 & Veilleux, S \\
M81 & 09:55:33.2 & 69:03:55.1 & -0.0001 & 2016 & Seth, A \\
M87 & 12:30:49.4 & 12:23:28.0 & 0.0043 & 2016 & Seth, A \\
M94 & 12:50:53.1 & 41:07:13.0 & 0.0010 & 2016 & Seth, A \\
M104 & 12:39:59.4 & -11:37:23.0 & 0.0036 & 2016 & Seth, A \\
MCG-05-23-016 & 09:47:40.1 & -30:56:56.0 & 0.0085 & 1670 & Shimizu, T \\
Mrk231 & 12:56:14.2 & 56:52:25.3 & 0.0422 & 1268 & Maiolino, R \\
Mrk273 & 13:44:42.1 & 55:53:13.5 & 0.0373 & 1717 & U, V \\
NGC0253 & 00:47:33.1 & -25:17:18.4 & 0.0008 & 1701 & Bolatto, A \\
NGC0424 & 01:11:27.5 & -38:05:01.8 & 0.0118 & 6138 & Marconcini, C \\
NGC1052 & 02:41:04.8 & -08:15:20.8 & 0.0052 & 2016 & Seth, A \\
NGC1068 & 02:42:40.8 & -00:00:45.9 & 0.0038 & 6138 & Marconcini, C \\
NGC1365 & 03:33:36.5 & -36:08:26.7 & 0.0055 & 6138 & Marconcini, C \\
NGC1566 & 04:20:00.2 & -54:56:17.2 & 0.0050 & 6138 & Marconcini, C \\
NGC1808 & 05:07:42.4 & -37:30:47.0 & 0.0033 & 6138 & Marconcini, C \\
NGC3081 & 09:59:29.5 & -22:49:34.8 & 0.0081 & 1670 & Shimizu, T \\
NGC3256N & 10:27:51.2 & -43:54:14.0 & 0.0094 & 1328 & Armus, L \\
NGC3256S & 10:27:51.2 & -43:54:19.2 & 0.0094 & 1328 & Armus, L \\
NGC4258 & 12:18:57.5 & 47:18:14.3 & 0.0015 & 2016 & Seth, A \\
NGC4395 & 12:25:48.9 & 33:32:48.7 & 0.0011 & 2016 & Seth, A \\
NGC5506 & 14:13:14.9 & -03:12:27.8 & 0.0061 & 1670 & Shimizu, T \\
NGC5728 & 14:42:23.9 & -17:15:11.1 & 0.0092 & 1670 & Shimizu, T \\
NGC6240 & 16:52:58.9 & 02:24:03.7 & 0.0243 & 1265 & Alonso-Herrero, A \\
NGC6552 & 18:00:07.3 & 66:36:54.3 & 0.0265 & 1039 & Dicken, D \\
NGC7172 & 22:02:01.9 & -31:52:10.5 & 0.0087 & 1670 & Shimizu, T \\
NGC7319 & 22:36:03.6 & 33:58:33.2 & 0.0225 & 2732 & Pontoppidan, K \\
NGC7469 & 23:03:15.6 & 08:52:26.0 & 0.0163 & 1328 & Armus, L \\
NGC7582 & 23:18:23.6 & -42:22:14.1 & 0.0054 & 3535 & Garcia-Bernete, I \\
UGC05101 & 09:35:51.7 & 61:21:12.3 & 0.0394 & 1717 & U, V \\
VV114 & 01:07:47.5 & -17:30:25.2 & 0.0201 & 1328 & Armus, L \\
VV340a & 14:57:00.7 & 24:37:02.8 & 0.0337 & 1717 & U, V \\
\hline
\end{tabular}
\tablefoot{The galaxy name, coordinates (RA and Dec), redshift, proposal ID (PID), and principal investigator (PI) of the MIRI/MRS proposal are provided, from left to right.}
\end{table*}

\section{[Mg\:V] flux distributions}

\begin{figure*}[!ht]
    \centering
 \includegraphics[width=0.32\textwidth]{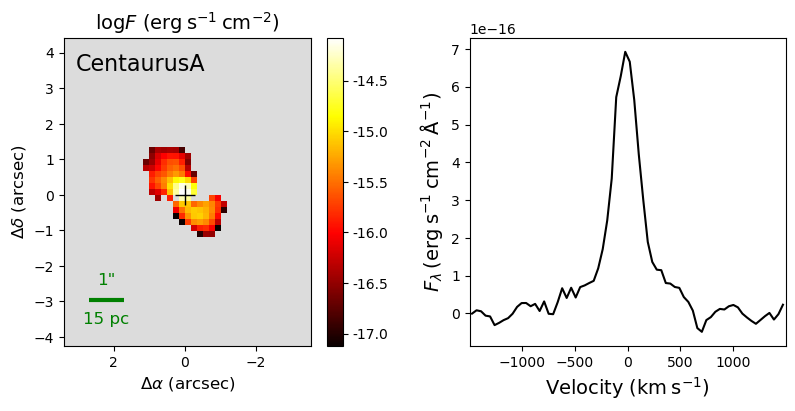}
  \includegraphics[width=0.32\textwidth]{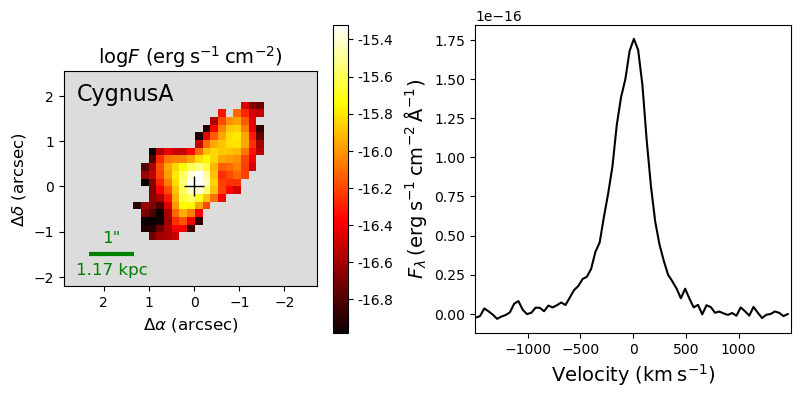}
  \includegraphics[width=0.32\textwidth]{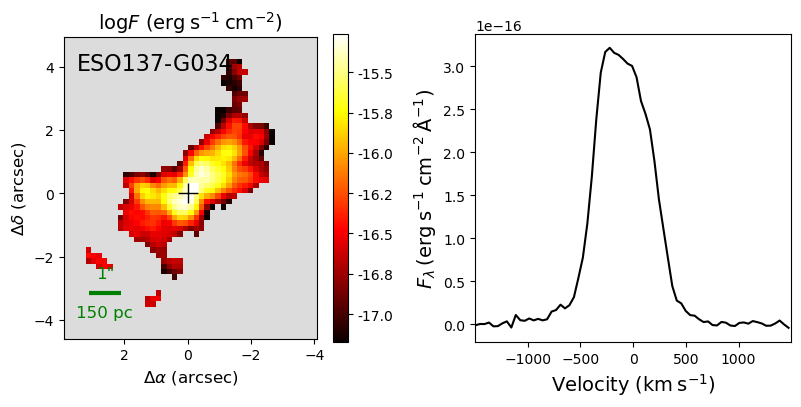}
 \includegraphics[width=0.32\textwidth]{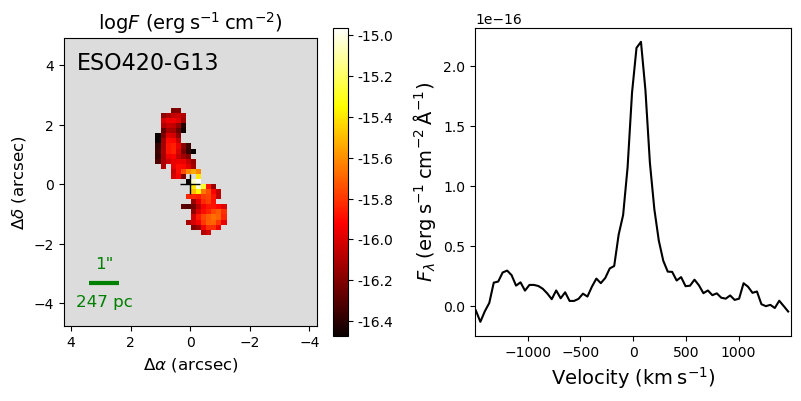}
 \includegraphics[width=0.32\textwidth]{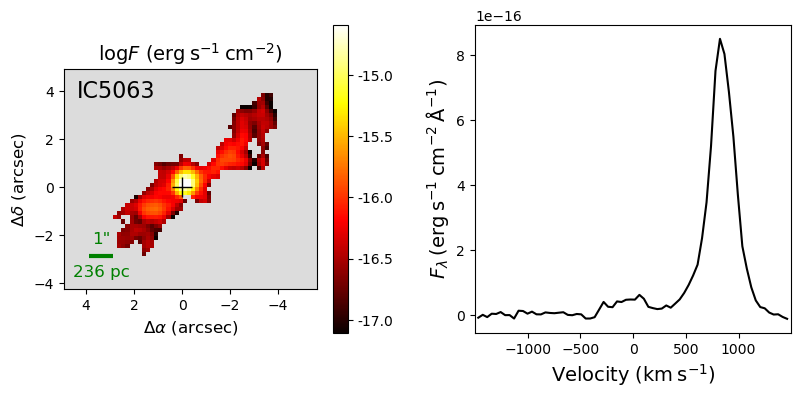}
 \includegraphics[width=0.32\textwidth]{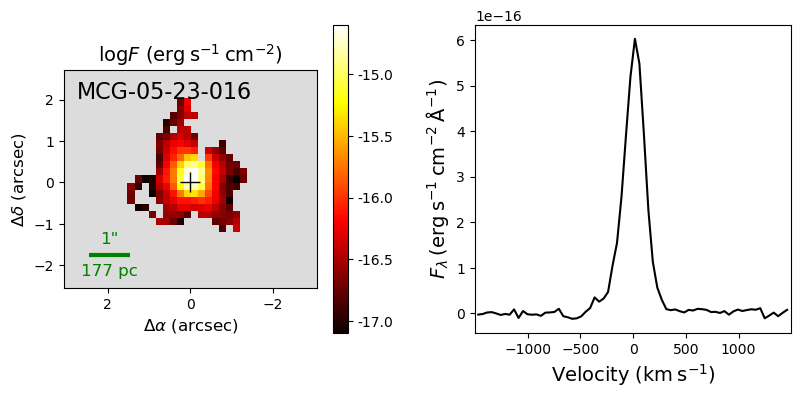}
 \includegraphics[width=0.32\textwidth]{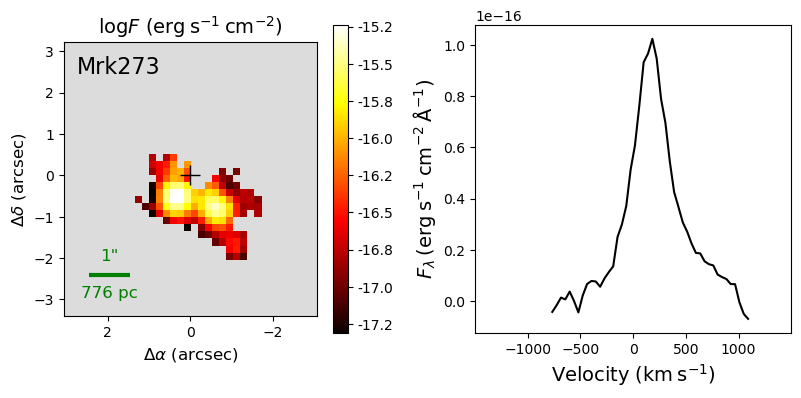}
 \includegraphics[width=0.32\textwidth]{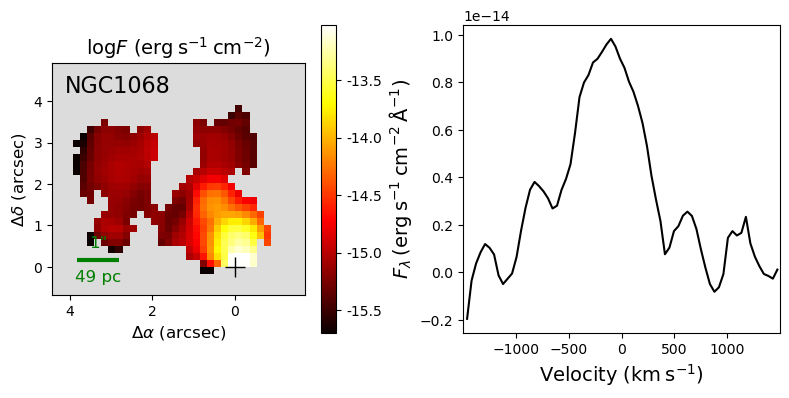}
 \includegraphics[width=0.32\textwidth]{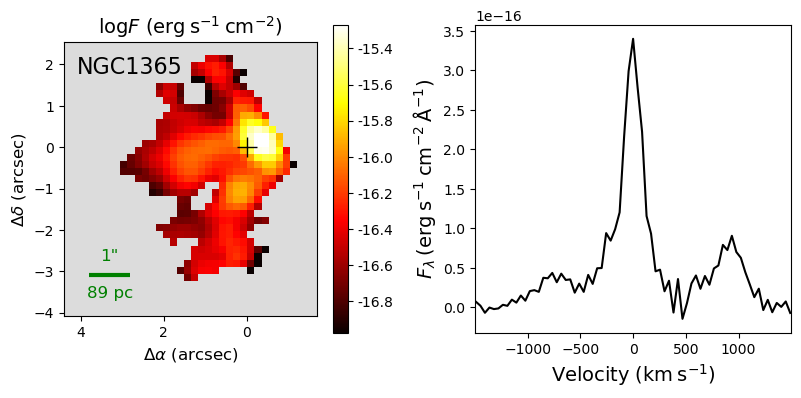}
 \includegraphics[width=0.32\textwidth]{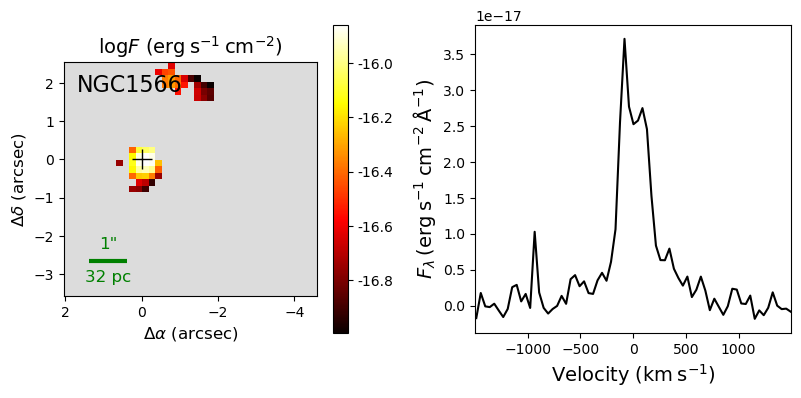}
 \includegraphics[width=0.32\textwidth]{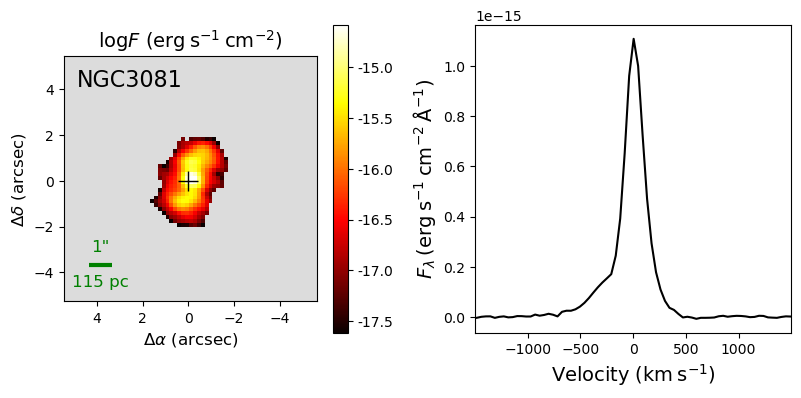}
 \includegraphics[width=0.32\textwidth]{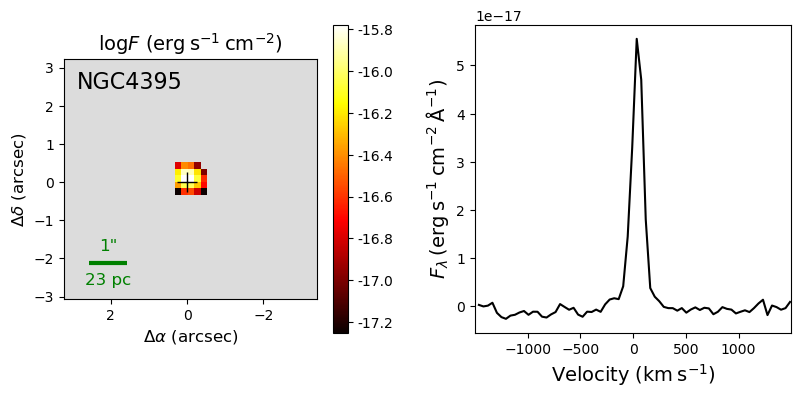}
 \includegraphics[width=0.32\textwidth]{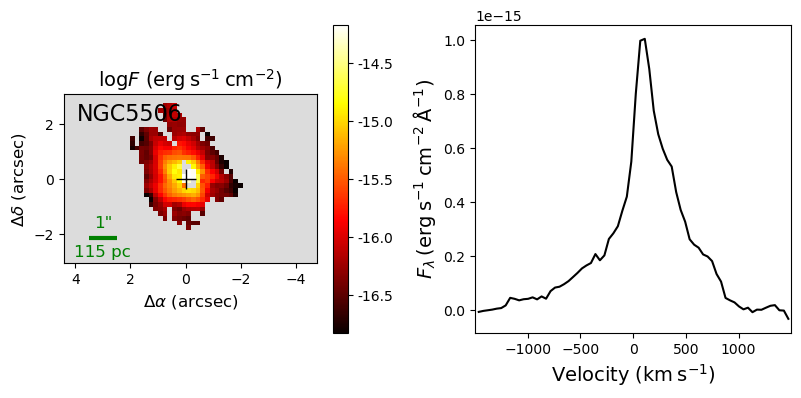}
 \includegraphics[width=0.32\textwidth]{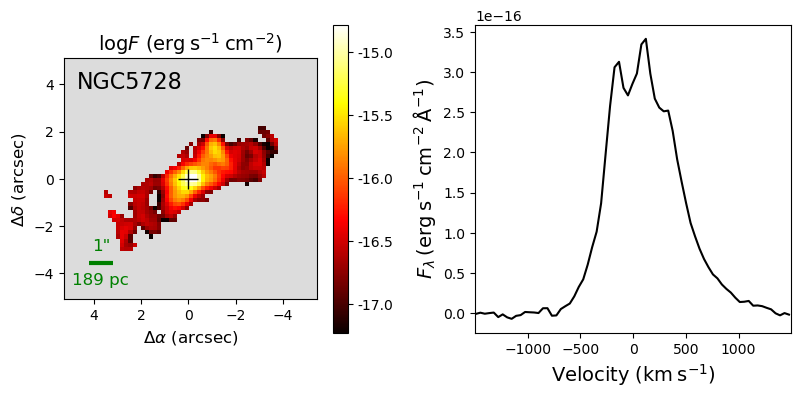}
  \includegraphics[width=0.32\textwidth]{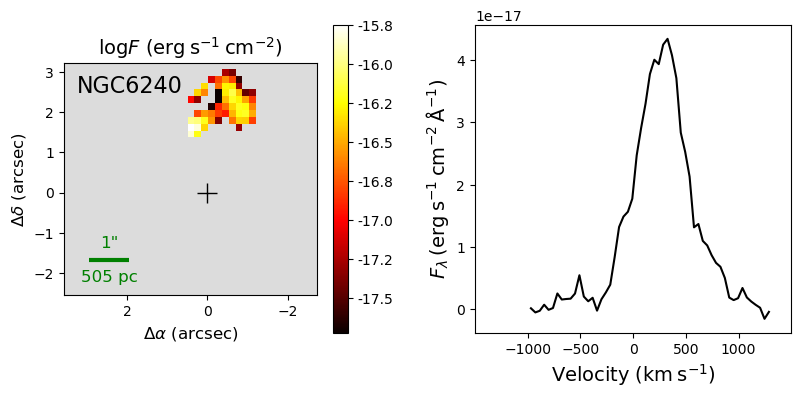}
 \includegraphics[width=0.32\textwidth]{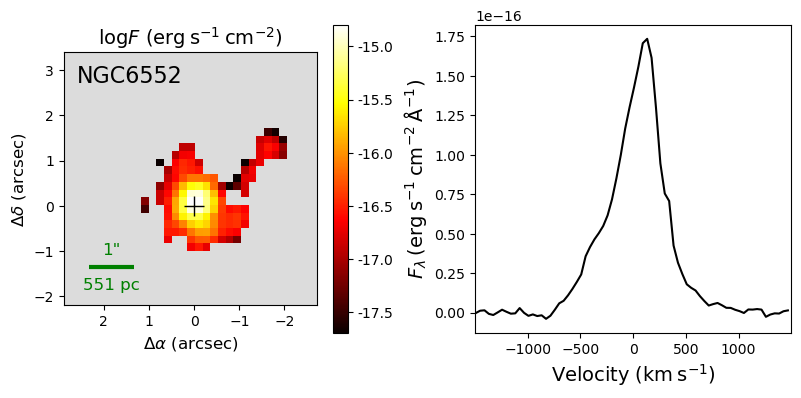}
 \includegraphics[width=0.32\textwidth]{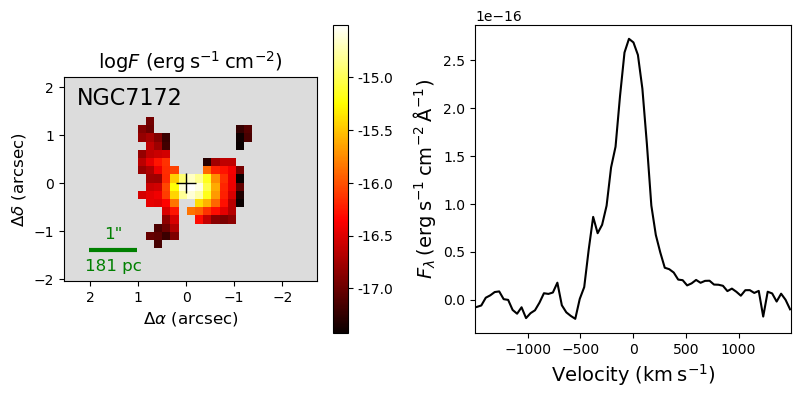}
 \includegraphics[width=0.32\textwidth]{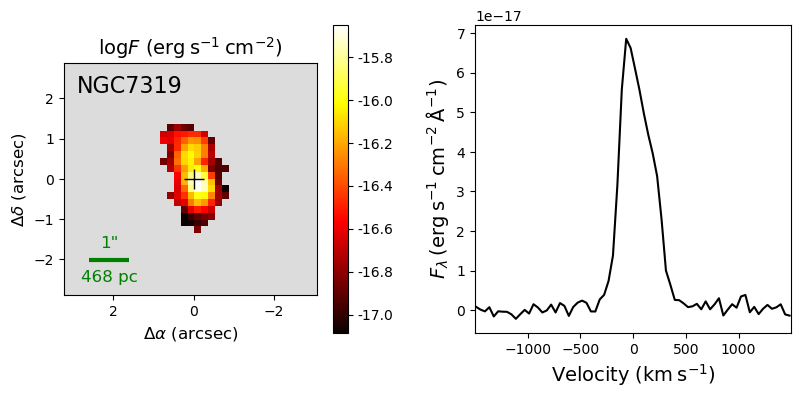}
 \includegraphics[width=0.32\textwidth]{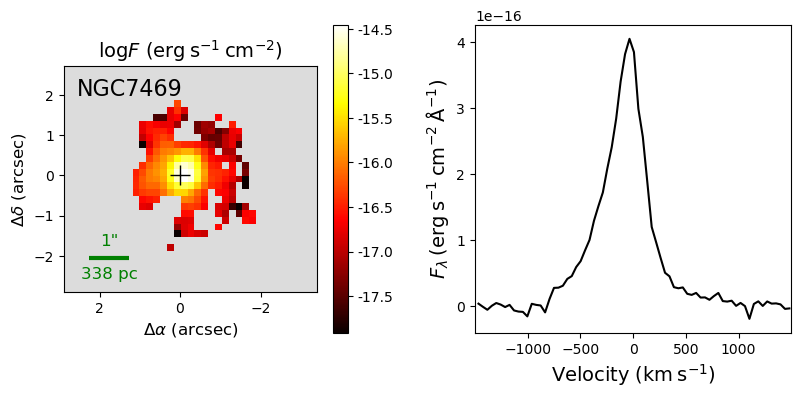}
 \includegraphics[width=0.32\textwidth]{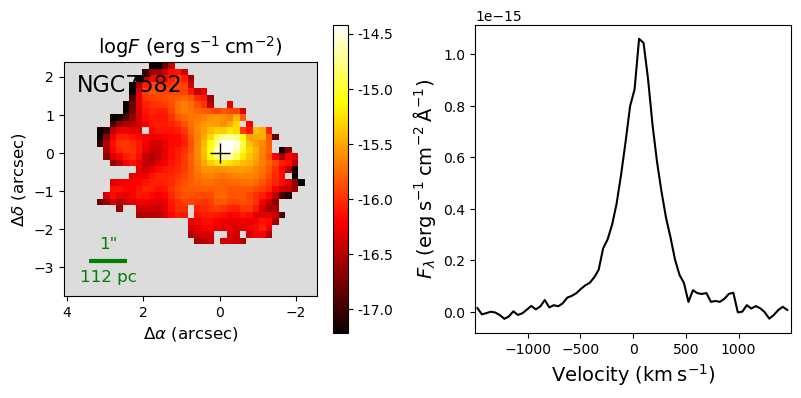}
 \includegraphics[width=0.32\textwidth]{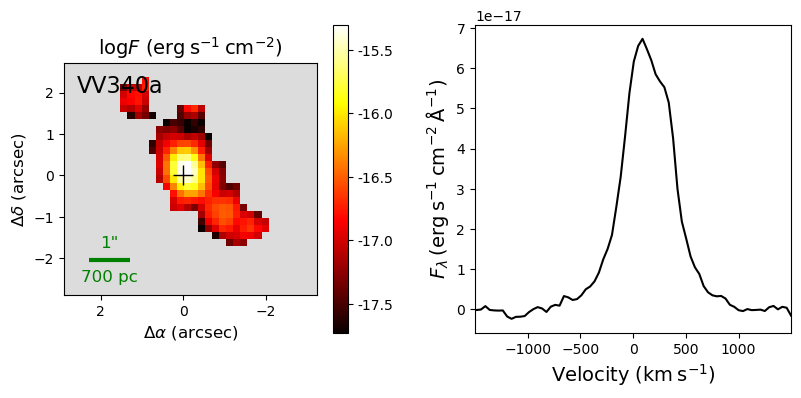}
  \caption{[Mg\:{\sc v}]$\lambda5.6098\mu$m flux distributions for galaxies with coronal line emission in our sample. The maps illustrate the flux distributions, with gray regions indicating areas where the corresponding emission line is not detected at $SNR > 5$. The plots display the [Mg\:{\sc v}]$\lambda5.6098\mu$m integrated profiles, obtained by summing the observed profiles across all detected spaxels and subtracting the continuum. The name of each galaxy is shown in the top-left corner of each plot, the spatial scale is shown in the bottom-left corner, and the central cross marks the position of the galaxy's nucleus.  }   
          \label{fig:mgv}
    \end{figure*}

In Fig.~\ref{fig:mgv} we present the flux maps for all galaxies with [Mg\:{\sc v}]$\lambda5.6098\mu$m emission detected in individual spaxels at a confidence level greater than 5 sigma, along with integrated line profiles considering all spaxels with detected emission. The sample galaxies exhibit diverse flux distributions, ranging from highly collimated structures to more rounded shapes. The discussion on the origin of coronal lines in this sample is beyond the scope of this paper,  as we use them solely as a selection criterion. 
Detailed results on some individual objects, including the analysis of coronal line emission, have already been published on the basis of MIRI MRS data \citep[e.g.][Veenema et al, in prep; among others]{Armus23,Hermosa24,HermosaMunoz25,Zhang24}.

\section{$W_{\rm 80}$ radial profiles per subsamples} \label{sec:W80_radial_types}

\begin{figure}
   \centering
   \includegraphics[width=0.49\textwidth]{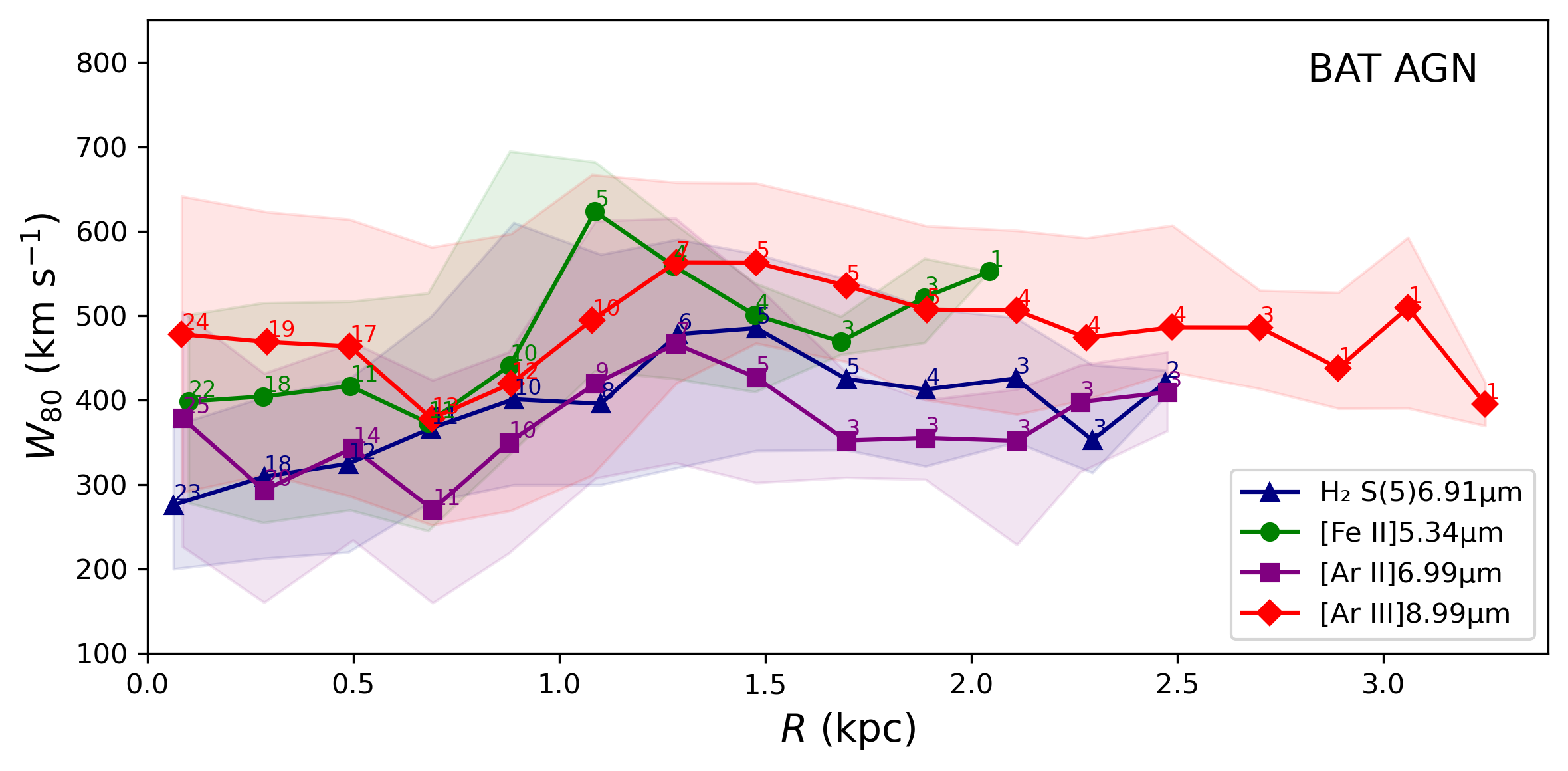}
     \includegraphics[width=0.49\textwidth]{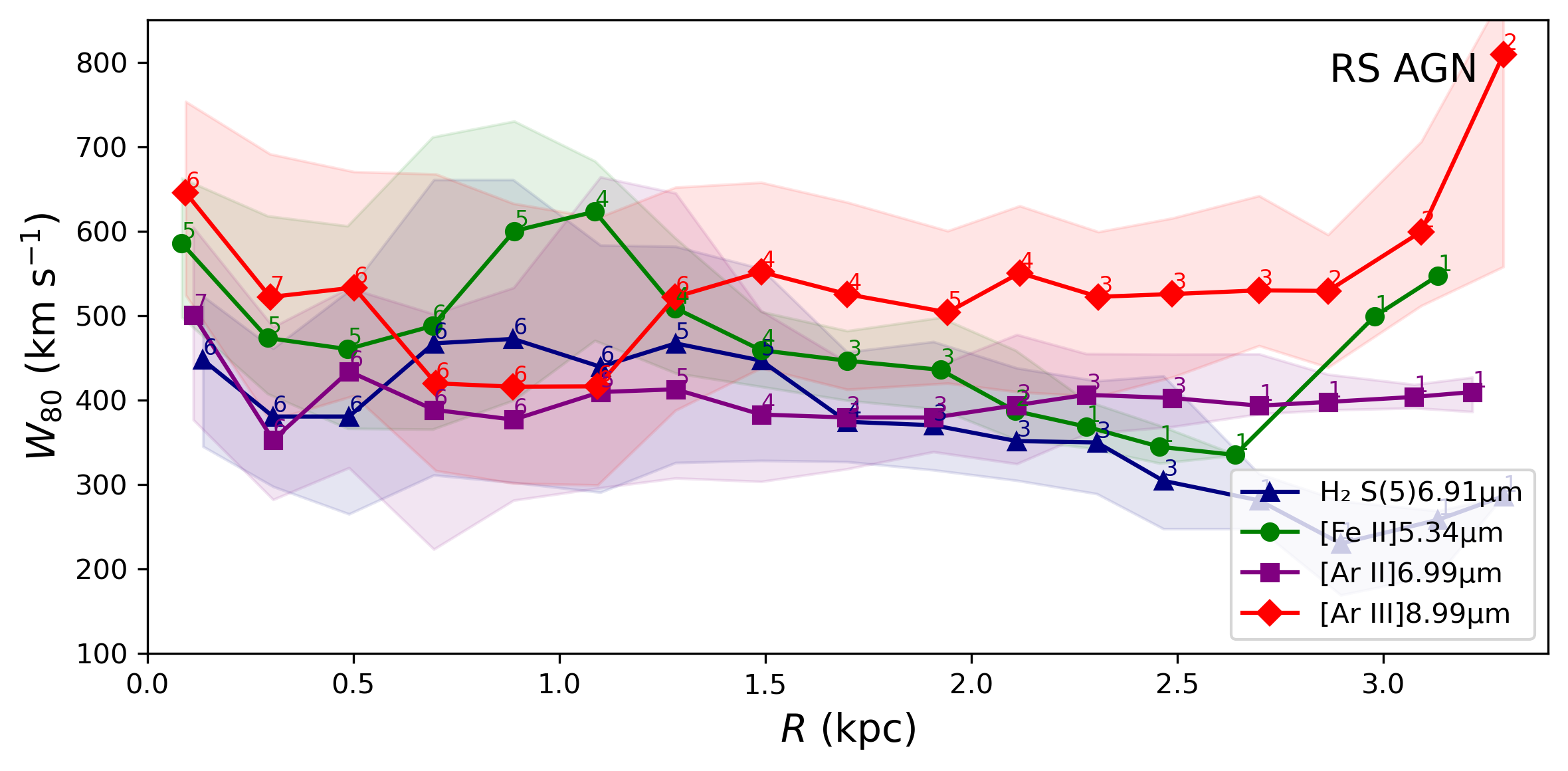}
     \includegraphics[width=0.49\textwidth]{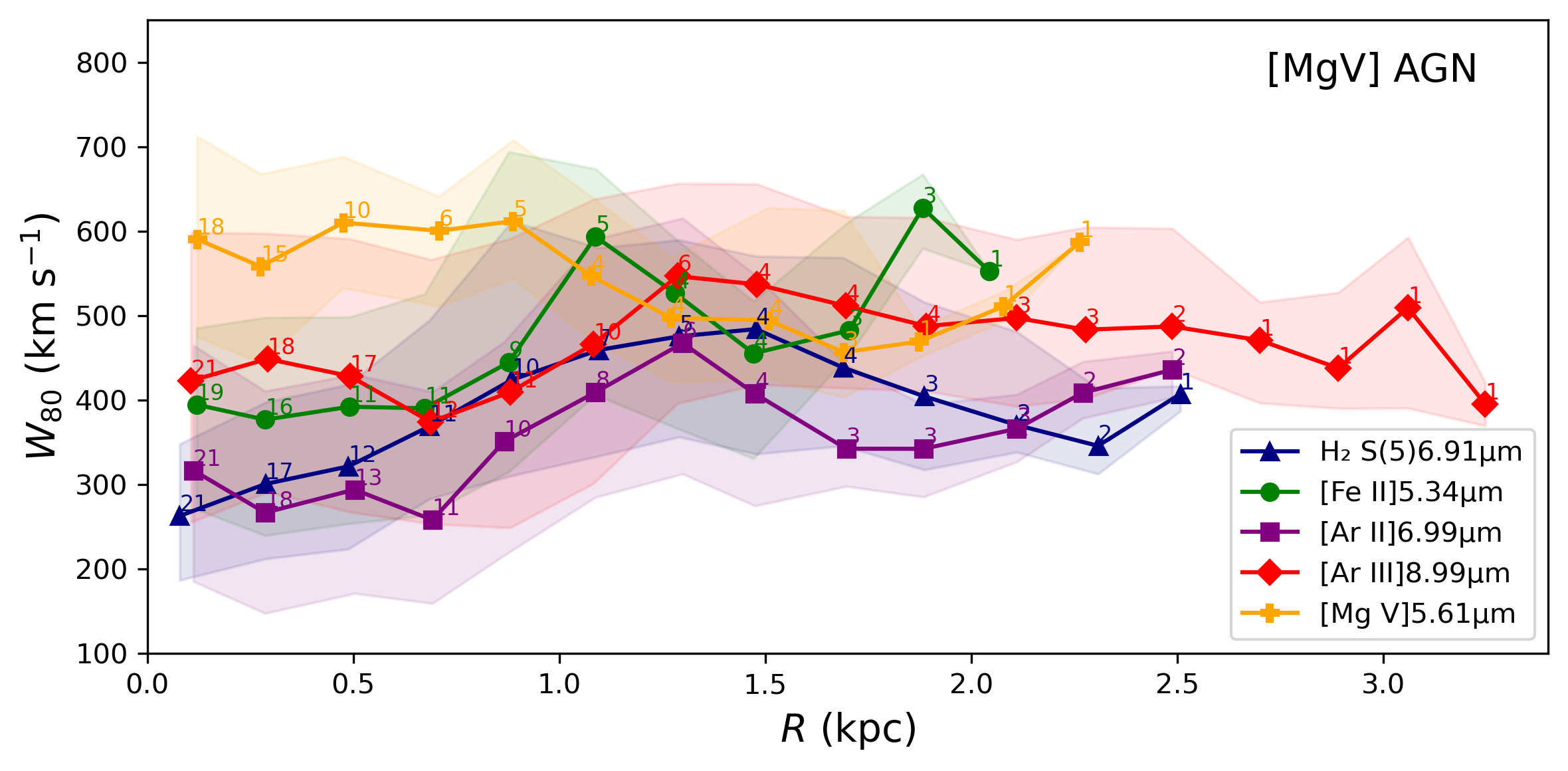}
     \includegraphics[width=0.49\textwidth]{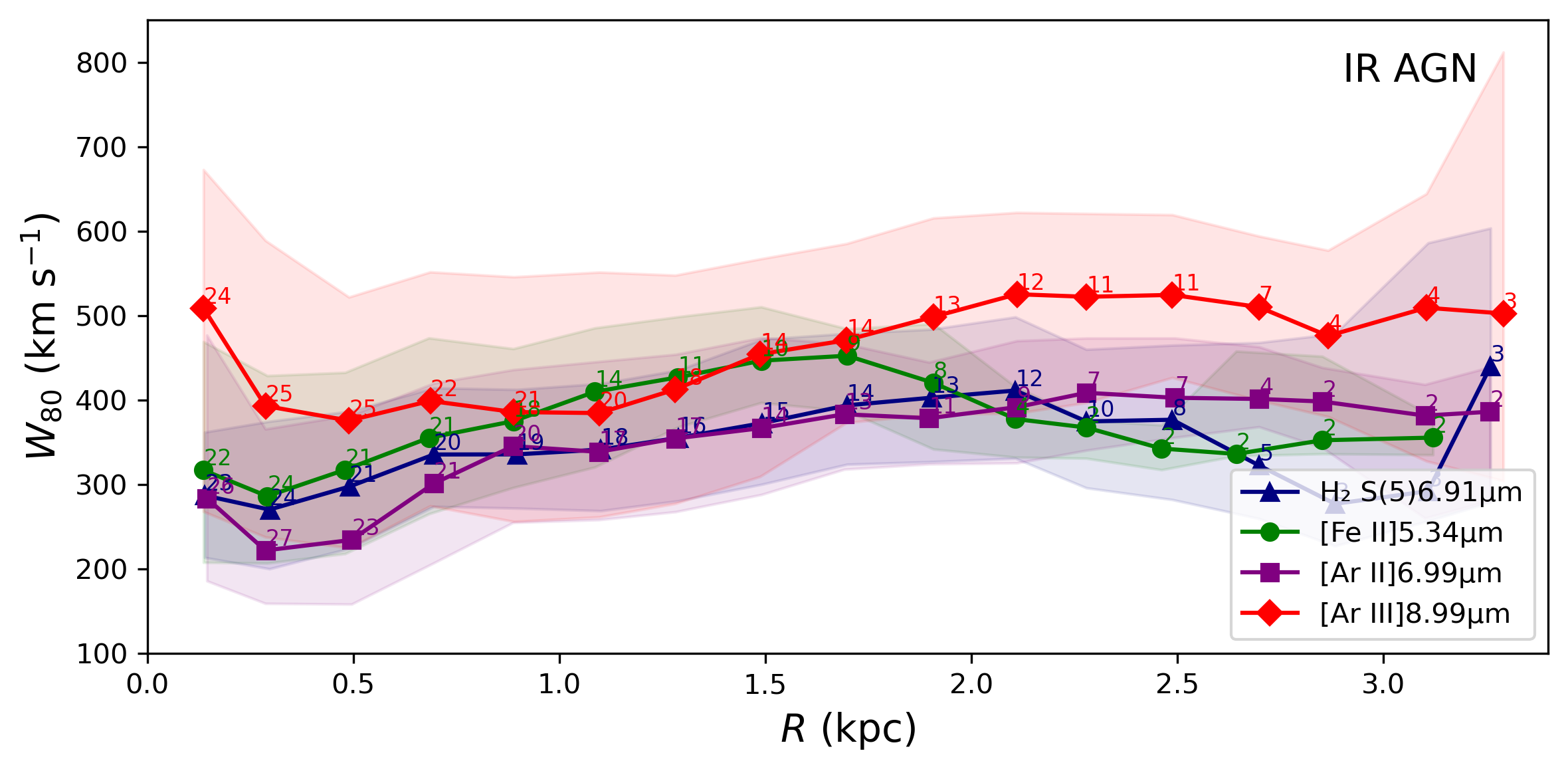}
      \includegraphics[width=0.49\textwidth]{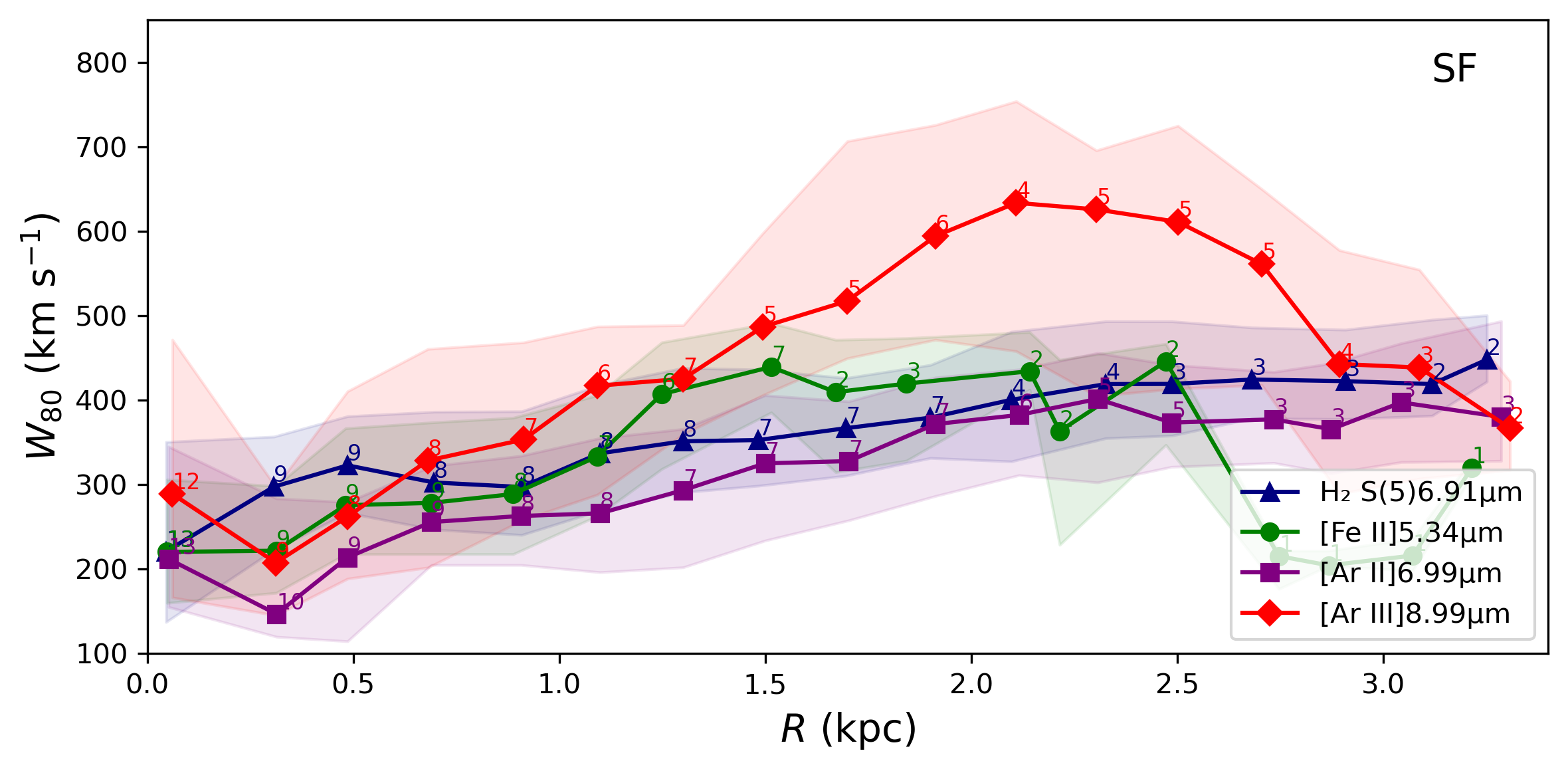}
     \caption{The $W_{\rm 80}$ radial profiles for the four emission lines are presented for the BAT AGN, RS AGN, [Mg\:{\sc v}] AGN, IR AGN and SF subsamples, from top to bottom, respectively. These profiles were computed following the same methodology as those shown in Fig.~\ref{fig:W80_radial_lines}.}              
         \label{fig:W80_radial_types}
   \end{figure}

 Fig.~\ref{fig:W80_radial_types} presents the $W_{\rm 80}$ radial profiles grouping the four emission lines for each subsample, providing a clearer comparison of the differences observed among the distinct emission lines. For the [Mg\:{\sc v}] AGN sample (bottom panel) we include also the [Mg\:{\sc v}]$\lambda5.6098\mu$m  $W_{\rm 80}$ values. For the SF sample, the nuclear $W_{\rm 80}$ values are typically below 300 km\:s$^{-1}$, reaching maximum values of $\sim$450 km\:s$^{-1}$ farther from the nucleus. The [Ar\:{\sc iii}] emission line is detected at radii larger than 1 kpc only for two objects in the SF sample (IRAS\:10565+2448 and IRAS\:23128-5919), which have higher $W_{\rm 80}$ values resulting in the increased values at these distances. For the AGN samples, the lowest $W_{\rm 80}$ values are typically observed for H$_2$ and [Ar\:{\sc ii}], which show similar values, while the highest values are found for [Fe\:{\sc ii}] and [Ar\:{\sc iii}].  Among AGN with coronal emission, within the inner $\sim$1 kpc, the highest $W_{\rm 80}$ values are observed for [Mg\:{\sc v}]. At larger distances, however, the coronal gas shows velocity dispersion similar to that of the other ionized gas lines.
 
\section{Shock and Photoionization models} \label{PhotModels}

\subsection{Shock models}

In this work, we use the fast shock models from \citet{Pereira-Santaella24b}. 
These models were computed with the MAPPINGS~V code \citep{Dopita96,Sutherland17} and are designed to investigate the origin of ionized gas emission lines in nearby galaxies observed with MIRI/MRS. 
They follow the prescription of \citet{Sutherland17}, varying several parameters: the gas metallicity ($Z_{\rm gas}$), the shock velocity ($v_s$), and defining a ram pressure variable as $R_{\rm P} = \frac{n_{\rm H}}{\text{cm}^{-3}} \times \left(\frac{v_s}{\text{km/s}}\right)^2$, where $n_{\rm H}$ is the gas volume density, and the magnetic-to-ram pressure ratio $\eta_{\rm M} = \frac{B^2}{4\pi\rho v_s^2}$, where $B$ is the magnetic field and $\rho$ the gas density.  In this work we adopt models with solar metallicity, shock velocities in the range 80–500~km\,s$^{-1}$, and ram pressure parameters $R_{\rm P} = 10^6$ and $10^8$. 
Further details on the shock models are given in \citet{Pereira-Santaella24b}.

\subsection{AGN and SF photoionization models}
We used \textsc{cloudy} code version c23.01 \citep{Ferland17,Chatzikos23} to construct grids of photoionization models for AGN and SF, following procedures similar to those described by \citet{Pereira-Santaella24b} and \citet{Dors12}.

For the AGN models, we adopt the ionizing Spectral Energy Distribution (SED) from \citet{Jin12}, based on the \textsc{optxagnf} model \citep{Done12}. This model describes accretion disk emission with three components powered by a single mass accretion flow: a multicolor blackbody disk producing optical/UV emission, a warm corona generating soft X-rays via Comptonization, and a hot corona responsible for hard X-rays. It is particularly suited for AGN with low Eddington ratios and is included in the \textsc{cloudy} SED library.

For the SF models, we generate the SEDs using the \textsc{Starburst99} code \citep{Leitherer99}, adopting an instantaneous SF mode. We assume a Kroupa Initial Mass Function (IMF) \citep{Kroupa02}, with a slope of $\alpha = 1.3$ for masses between 0.1 and 0.5\,M$_{\odot}$, and $\alpha = 2.3$ for masses from 0.5 to 100\,M$_{\odot}$. We use the Padova evolutionary tracks with solar metallicity \citep{Bressan93}. The geometry is assumed to be plane-parallel, and the ionization parameter ($U$) is used to scale the source intensity. We used the SEDs output by the \textsc{Starburst99} code for ages ranging from 1 to 10 Myr, in 1 Myr increments, to construct the photoionization models with \textsc{cloudy}.

For both AGN and SF models, we adopt solar abundances for all elements based on \citet{Grevesse10}, with an oxygen abundance of $12 + \log({\rm O/H}) = 8.69$. The only exception is iron, which is known to exhibit significant scatter at a fixed O/H value \citep{Izotov06}. Therefore, we estimate the iron abundance using the relation between iron and oxygen abundances presented by \citet{Izotov06}, resulting in $\log({\rm Fe/H})\approx-5.60$ for solar oxygen abundance, which is adopted in our models.  We adopt grain abundances based on the Orion Nebula, but scale the PAH abundance to match the range observed in local galaxies, as reported by \citet{Draine07}.

We computed a sequence of models with densities spanning $2 \leq \log n_{\rm H}/{\rm cm^{-3}} \leq 6$ and ionization parameters covering $-4.0 \leq \log U \leq -1.0$, both sampled in increments of 1 dex. The models were stopped once the gas temperature fell below 4000\,K, the default stopping criterion in \textsc{Cloudy}.  This implies that our models do not predict the H$_2$ emission lines, which are produced at lower temperatures. The stopping criterion is necessary to obtain robust predictions for the ionized gas emission lines, avoiding the complex physics of the transition regions to neutral and molecular gas. Moreover, \textsc{Cloudy} assumes a simplified one-dimensional structure for AGN, with a central point-like ionizing source and gas in static layers, which is sufficient for modeling the ionized gas emission but does not capture the detailed three-dimensional geometry or the physical conditions in the colder molecular gas regions.

\end{document}